\documentclass[12pt,document,nofootinbib,superscriptaddress, onecolumn,preprintnumbers,balancelastpage]{revtex4}
\pdfoutput=1
\usepackage{latexsym}
\usepackage{amssymb}
\usepackage{epsfig,amsmath,graphics}
\usepackage{color}
\usepackage{dcolumn}
\usepackage{slashed}
\usepackage{comment,latexsym} 
\usepackage{bm}
\usepackage{verbatim}
\usepackage{tabularx}
\usepackage{dcolumn}
\usepackage{hyperref}
\usepackage{setspace}
\usepackage{xspace}
\usepackage{bigints}

\newcommand{\OO}{\mathcal{O}}

\def\be{\begin{equation}}
\def\ee{\end{equation}}
\newcommand{\beq}{\begin{equation}}
\newcommand{\eeq}{\end{equation}}
\def\bea{\begin{eqnarray}}
\def\eea{\end{eqnarray}}
\newcommand{\eref}[1]{Eq.~(\ref{#1})}

\newcommand{\MeV}{{\text{ MeV}}}
\newcommand{\GeV}{{\text{ GeV}}}
\newcommand{\TeV}{{\text{ TeV}}}

\renewcommand{\deg}{\ensuremath{^{\circ}}\xspace}

\begin{document}

\begin{flushright}
\text{\normalsize MCTP-13-19} \\
\vspace{-5pt}
\text{\normalsize MIT-CTP 4482}\\
\vspace{-5pt}
\text{\normalsize SLAC-PUB-15664}
\end{flushright}
\vskip 45 pt

\title{Wino Dark Matter Under Siege}

\author{Timothy Cohen}
\affiliation{
Theory Group, SLAC National Accelerator Laboratory\\
\vskip -6 pt
Menlo Park, CA, 94025}

\author{Mariangela Lisanti}
\affiliation{
Princeton Center for Theoretical Science, Princeton University\\
\vskip -6 pt
Princeton, NJ 08544}

\author{Aaron Pierce}
\affiliation{
Michigan Center for Theoretical Physics, Department of Physics\\
\vskip -6 pt
Ann Arbor, MI 48109}

\author{Tracy R. Slatyer}
\affiliation{
School of Natural Sciences, Institute for Advanced Study\\
\vskip -6 pt
Princeton, NJ 08540}
\affiliation{
Center for Theoretical Physics, Massachusetts Institute of Technology\\
\vskip -6 pt
Cambridge, MA 02139}

\begin{abstract}
\vskip 15 pt
\begin{center}
{\bf Abstract}
\end{center}
\vskip -30 pt
$\quad$
\begin{spacing}{1.2}
A fermion triplet of $SU(2)_L$ -- a wino -- is  a well-motivated dark matter candidate.  This work shows that present-day wino annihilations are constrained by indirect detection experiments, with the strongest limits coming from H.E.S.S. and Fermi.  The bounds on wino dark matter are presented as a function of mass for two scenarios: thermal (winos constitute a subdominant component of the dark matter for masses less than 3.1 TeV) and non-thermal (winos comprise all the dark matter).  Assuming the NFW halo model, the H.E.S.S. search for gamma-ray lines excludes the 3.1 TeV thermal wino; the combined H.E.S.S. and Fermi results completely exclude the non-thermal scenario.  Uncertainties in the exclusions are explored.  Indirect detection may provide the only probe for models of anomaly plus gravity mediation where the wino is the lightest superpartner and scalars reside at the 100 TeV scale.
\end{spacing}
\end{abstract}


\maketitle
\newpage
\section{Introduction}

A weakly interacting massive particle (WIMP) is a well-motivated candidate for the Universe's missing matter.
Direct detection experiments, however, continue to tighten limits on $\OO(100 \GeV)$ mass WIMPs.  
Furthermore, the 8 TeV Large Hadron Collider (LHC) has found no evidence for the lightest stable neutral particle of supersymmetry, which is often associated with the WIMP.  These null results might suggest that the dark matter (DM) is still a WIMP, but with a somewhat heavier mass in the multi-TeV range.  It is crucially important to understand the limits on TeV-scale WIMPs.

A TeV-scale WIMP candidate that has an annihilation cross section consistent with that of a thermal relic is the $SU(2)_L$ triplet fermion, $\chi$.  A minimal model with the multiplet $\chi$ added to the Standard Model is
\begin{equation}
\mathcal{L} = \mathcal{L}_\mathrm{SM} + \bar{\chi} \left( i\slashed{D} + M_2\right)\chi,
\end{equation}
which has a single free parameter, the mass $M_{2}$, and has been explored in detail in~\cite{Hisano:2002fk, Hisano:2003ec, Hisano:2004ds, Hisano:2006nn, Cirelli:2005uq, Cirelli:2007xd, Ciafaloni:2010ti, Hryczuk:2011vi}.  The state $\chi$ has the same quantum numbers as the superpartner of the weak gauge bosons and, borrowing the terminology from the minimal supersymmetric standard model embedding, we refer to it as wino DM.  Throughout this paper, the assumption is that the wino is ``pure" and has (approximately) no mixing with other neutralinos.  

The wino is the lightest superpartner in a variety of models -- for example, theories where anomaly-mediated supersymmetry breaking determines the gaugino masses~\cite{Giudice:1998xp,Randall:1998uk}.  If the gravity-mediated contribution to scalars is unsequestered, the scalars are a loop-factor heavier than the gauginos.  The characteristic spectra of these ``split supersymmetry'' models~\cite{Wells:2004di, ArkaniHamed:2004fb,Giudice:2004tc}  have weak-scale gluinos and neutralinos, with all other superpartners out of reach for current experiments.  Split supersymmetry has drawn renewed interest in light of the Higgs boson mass measurement and the absence of other direct evidence for superpartners~\cite{Arvanitaki:2012ps,ArkaniHamed:2012gw,LawrenceYasunori, Kane:2011kj, Ibe:2011aa}.  Because the heavy scalars in these models apparently point to fine-tuning, naturalness can no longer be invoked to anchor the lightest superpartner to the weak scale.  However, if this state accounts for the relic density of the DM, the ``WIMP miracle'' indicates that it should not be too much heavier than the $W^\pm$ boson~\cite{ArkaniHamed:2004fb, Giudice:2004tc,Pierce:2004mk,Feng:2011ik}.

Given the Planck measurement $\Omega \,h^2 = 0.1199 \pm 0.0027$ \cite{Ade:2013zuv}, a $\sim 3.1 \TeV$ thermal wino can comprise all of the DM. 
This DM candidate is difficult to observe at any foreseeable collider.   Additionally, because the wino has no renormalizable interactions with the Higgs boson, its tree-level spin-independent scattering cross section with nucleons is zero; loop diagrams yield an observable signal well below the current bounds from direct detection, but just above the neutrino floor \cite{Hisano:2010fy, Hill:2011be}.  Our purpose here is to challenge the pessimism associated with testing the thermal wino hypothesis.  
When the wino mass is significantly larger than the $W^{\pm}$-boson mass, the non-perturbative effect known as the Sommerfeld enhancement (SE), which becomes large at low velocities, substantially enhances the annihilation cross section of winos in the Universe today \cite{Hisano:2003ec, Hisano:2004ds}.  An observable number of photons results and existing gamma-ray telescopes are sensitive to a large fraction of the interesting parameter space. In this work the current status of the experimental limits on wino DM is explored.  A complementary paper \cite{MattJiJi} also studies the implications of these limits, especially with regards to non-thermal scenarios.

The rest of this paper is organized as follows.  Sec.~\ref{sec:constraining} reviews the current bounds on the wino, with an emphasis on the status of indirect detection experiments.  Sec.~\ref{sec:uncertainties} discusses some of the astrophysical and theoretical uncertainties and Sec.~\ref{sec:projections} presents future projections for wino detection.  Appendix \ref{Appendix: Sommerfelding} provides the technical details needed to accurately compute the SE, and Appendix~\ref{Appendix: Subtraction} reviews some more detailed aspects of the one-loop-SE calculation.

\section{Constraining Winos}
\label{sec:constraining}
 
As mentioned above, the current LHC and direct detection measurements do not strongly constrain wino DM.  However, indirect detection constraints from the Fermi Gamma-Ray Space Telescope (``Fermi'') and the High Energy Spectroscopic System (``H.E.S.S.'') are highly relevant.  This section presents the bounds for two cosmological scenarios: a thermal cosmology, where the relic abundance is equal to its thermal freeze-out value, and a non-thermal cosmology, where the relic abundance is set equal to the measured Planck value by some unspecified dynamics in the early Universe, \emph{e.g.}~the late decay of a modulus \cite{Moroi:1999zb, Gelmini:2006pw, Acharya:2009zt,Moroi:2013sla, Easther:2013nga}.

A wino multiplet consists of a neutral Majorana fermion (the neutralino $\chi^0$) and a charged fermion (the chargino $\chi^\pm$).  A radiative mass splitting $\delta$ between these states is induced at one-loop by the exchange of electroweak gauge bosons.  The mass splitting is an effect of electroweak symmetry breaking, so its value is calculable in the effective theory and is cut-off by the weak scale.  In the pure-wino limit, the mass splitting to two-loop accuracy is\footnote{For the plots in this paper, $\delta$ is set to $0.17 \GeV$ independent of energy.
The exact value of $\delta$ determines the position of the resonance and therefore has the largest effect near 2.3 TeV.  
}
\be\label{Eq: delta}
\delta = 0.1645\pm0.0004 \GeV 
\ee
for $M_2 = 2 \TeV$.  There is a relatively mild dependence on $M_2$: $\delta \simeq 150 \MeV$ at 100 GeV and asymptotes to \eref{Eq: delta} for wino masses above 1\TeV~\cite{Ibe:2012sx}.  In supersymmetric models, mixing between the wino and other neutralinos may modify this splitting.  However, the leading operator that splits the charged and neutral wino is dimension seven:
\begin{equation}
{\mathcal O}_\delta \sim  \chi^{a} \chi^{b} \left(H^{\dagger} T^{a} H \right)  \left(H^{\dagger} T^{b} H\right),
\end{equation}
where $\chi^a$ is the full wino multiplet and $H$ is the Higgs field.  Given its high dimension, this operator quickly decouples as the Higgsino mass $\mu$ and $M_{1,2}$ rise above the weak scale, implying that \eref{Eq: delta} holds for a large class of models.  The approximate degeneracy of the charged and neutral states has important observational consequences.

At the LHC, wino-like charginos can be directly produced.  The small mass splitting allows the decay $\chi^\pm \rightarrow \chi^0\,\pi^\pm$ with a lifetime that is $\mathcal{O}(10 \text{ cm})$; the pion produced in the decay is typically too soft to observe, and the event can only be characterized by a disappearing charged track.  A 7 TeV LHC search for this signature \cite{ATLAS:2012jp} places a lower limit of $\sim108 \GeV$ on the wino mass~\cite{Ibe:2012sx}.  It is also possible to search for directly produced wino-like neutralinos, simply by looking for missing energy plus a jet from initial-state radiation.  Current monojet searches at the LHC do not constrain the pure-wino limit~\cite{ATLAS-CONF-2012-147, CMS-PAS-EXO-12-048}.

Direct detection limits on the pure-wino scenario are currently non-existent.  Because there is no tree-level wino-wino-Higgs coupling, the elastic scattering of a wino off a nucleon occurs at one-loop (coupling to quarks in the nucleon) and two-loop (coupling to gluons).  The associated spin-independent cross section is $\mathcal{O}\big(10^{-47} \text{ cm$^2$}\big)$ for $50\GeV$ to $3\TeV$ winos~\cite{Hisano:2011cs}, which is well below the strongest direct detection limits to date (currently from the Xenon100 experiment~\cite{Aprile:2012nq}).
\begin{figure}[t!]
\centering
\vspace{-50 pt}
\includegraphics[width=0.7\textwidth]{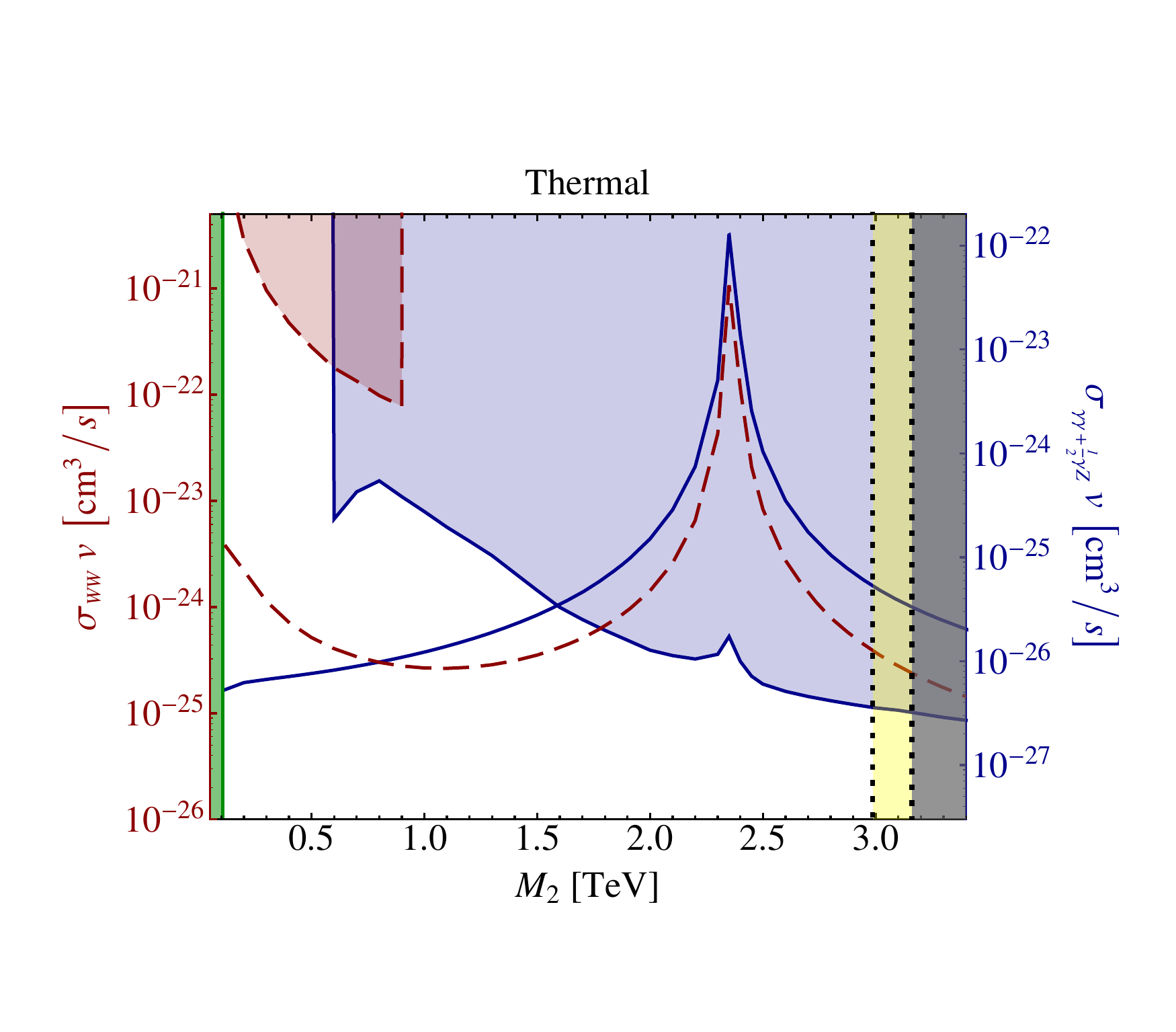}
\vspace{-35pt}
\caption{
The dashed red line shows $\sigma\big(\chi^0\,\chi^0 \rightarrow W^+\,W^-\big)\,v$ in cm$^3$/s.  The solid blue line shows $\sigma\big(\chi^0\,\chi^0 \rightarrow \gamma\,\gamma\big)\,v + \frac{1}{2}\sigma\big(\chi^0\,\chi^0 \rightarrow \gamma \,Z^0\big)\,v$ in cm$^3$/s.  All three cross sections are computed in the tree-level-SE approximation. \emph{One-loop effects have been shown to reduce the cross section to line photons by as much as a factor of 4} (see Sec.~\ref{Sec: 1-Loop}).  
The exclusion from Fermi (relevant for the $W^+\,W^- $ channel) is the shaded red region, which is bordered by the dashed line.  The exclusion from H.E.S.S. (relevant for the $ \gamma\,\gamma + \frac{1}{2} \gamma \,Z^0$ channel) is the shaded blue region, which is bordered by the solid line.  These exclusion contours assume that the wino abundance is set by thermal freeze-out. The H.E.S.S. limit is appropriate for an NFW profile, see Sec.~\ref{sec: Astro}.  The shaded yellow region between the dotted lines corresponds to $\Omega\, h^2 = 0.12\pm 0.006$.  In the black shaded region, a thermal wino exceeds the observed relic density.
}
\label{Fig: Wino Exclusion Thermal}
\end{figure}

Indirect detection experiments can cover the broad region of wino parameter space to which the LHC and direct detection experiments are not sensitive.  In particular, if the wino makes up a non-trivial fraction of the DM, it can lead to observable rates for experiments that search for photons from DM annihilation.  Even in this case, the perturbative annihilation cross section for winos is not always large enough to be observable.  However, as the wino mass becomes large with respect to the $W^{\pm}$-boson mass, non-perturbative SE effects due to the presence of a relatively long-range potential become important, especially at low velocities.  The impact of the SE on wino annihilation has been studied in detail~\cite{Hisano:2002fk, Hisano:2003ec, Hisano:2004ds, Hisano:2006nn, Cirelli:2005uq, Cirelli:2007xd, Ciafaloni:2010ti, Hryczuk:2011vi} and must be properly accounted for when computing the wino relic density, as well as its present-day annihilation cross section.  Following \cite{Hisano:2002fk, Hisano:2003ec, Hisano:2004ds, Hisano:2006nn}, we take the mass dependence for most cross sections to be proportional to $1/M_2^2$.  However, we include the appropriate phase-space and propagator factors for wino annihilations to $W^+ W^-$ and $\gamma\,Z^0$ today as they are numerically relevant at low mass.  This implies that our relic density is a slight overestimate at $\OO(100 \GeV)$ masses.  Appendix~\ref{Appendix: Sommerfelding} reviews the procedure we follow to compute these non-perturbative effects, and we refer the reader there for an overview of the computation, as well as a description of the procedure used to minimize numerical convergence problems.  
 
A number of ground- \cite{Abramowski:2013ax, Abramowski:2010aa, AlexGeringer-SamethfortheVERITAS:2013fra, Aliu:2012ga, Aleksic:2011jx} and space-based~\cite{Adriani:2010rc, Ackermann:2012qk, Ackermann:2011wa} experiments place significant constraints on wino annihilation.  The strongest and most robust bounds come from Fermi \cite{Ackermann:2011wa}, for $100 \GeV \lesssim M_2 \lesssim 900 \GeV$, and H.E.S.S. \cite{Abramowski:2013ax}, for $600 \GeV \lesssim M_2 \lesssim  25 \TeV$.  The Fermi result is derived by stacking 24 months of data for ten satellite galaxies and places limits on the continuum photons from DM annihilation to $W^+W^-$. The Fermi collaboration has recently presented updated limits from fifteen dwarf galaxies that are weaker by a factor of $\sim 2$ \cite{FermiTalk}; in this work, we use the published bound \cite{Ackermann:2011wa}.  

The published Fermi limit on annihilation to $W^+W^-$ is roughly comparable to that obtained from the antiproton flux measurement by PAMELA~\cite{Adriani:2010rc, Belanger:2012ta, Cirelli:2013hv}.  The antiproton measurement is subject to uncertainties from the DM profile, as well as the antiproton propagation parameters.  The choice of the propagation model can cause one or two orders of magnitude uncertainty in the limits~\cite{Belanger:2012ta, Cirelli:2013hv,Evoli:2011id}.  For this reason, the PAMELA antiproton limits will not be explored further.  The positron excess observed by PAMELA~\cite{Adriani:2008zr} and Fermi~\cite{FermiLAT:2011ab}, and recently confirmed to high precision by AMS-02~\cite{Aguilar:2013}, has smaller errors associated with the positron propagation, but the resulting bound is several orders of magnitude weaker than the antiproton and dwarf gamma-ray constraints for $W^+W^-$ annihilation~\cite{Kopp:2013eka}.

\begin{figure}[t!]
\centering
\vspace{-50pt}
\includegraphics[width=0.7\textwidth]{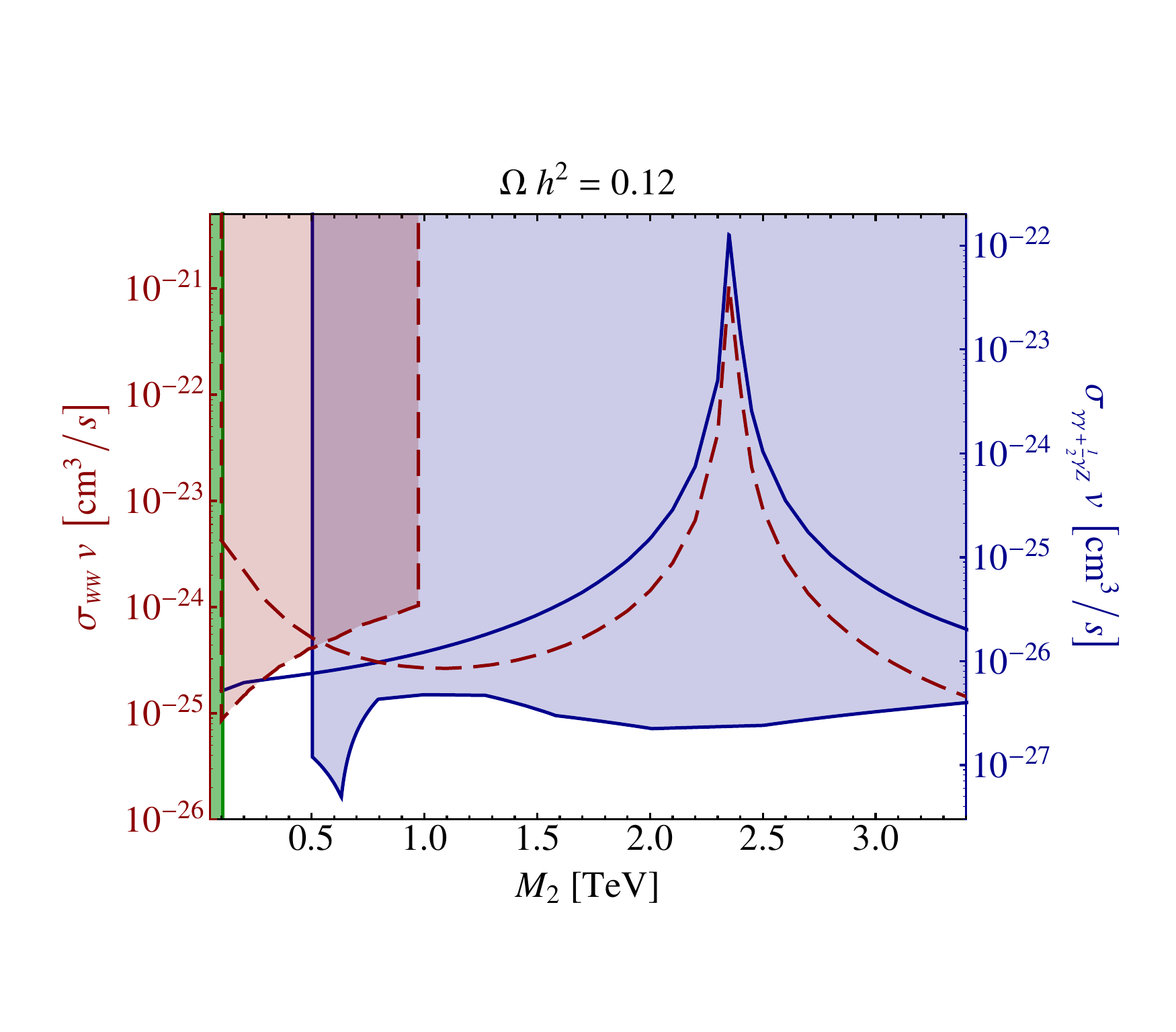} 
\vspace{-35pt}
\caption{
As in Fig.~\ref{Fig: Wino Exclusion Thermal}, but now exclusion contours assume the wino comprises all the DM as the result of an unspecified non-thermal history.  
}
\label{Fig: Wino Exclusion Non Thermal}
\end{figure}

The H.E.S.S. limit arises from a search for gamma-ray lines in a $1\deg$ radius circle at the Galactic Center, with the Galactic plane excluded by restricting the Galactic latitude to $|b| > 0.3 \deg$.  An earlier H.E.S.S. analysis searched for continuum gamma-rays from the Galactic Center \cite{Abramowski:2011hc} and relied on a spatial subtraction of the background.  No bound can be placed using this procedure if a DM core extends beyond a radius of $\OO(0.1\,\text{kpc})$.  Moreover, the bounds are generally weaker (see \cite{Cirelli:2007xd}) than the line search considered here, even for the NFW profile.\footnote{For example, assuming an NFW profile, a 3 TeV wino that comprises all the DM would escape detection by a factor of $\sim 2$ in the spatial subtraction analysis of Ref.~\cite{Abramowski:2011hc}, while it is excluded by a factor of $\sim 15$ (for tree-level-SE) in the line analysis (see Fig.~\ref{Fig: Wino Exclusion Non Thermal}).}  Therefore, we concentrate on the line search.

Figs.~\ref{Fig: Wino Exclusion Thermal} and \ref{Fig: Wino Exclusion Non Thermal} summarize the limits on the pure-wino scenario.  Fig.~\ref{Fig: Wino Exclusion Thermal} applies when the wino's relic density is  equal to its thermal abundance.  Fig.~\ref{Fig: Wino Exclusion Non Thermal} assumes that the wino relic density is equal to the measured value, requiring an unspecified non-thermal cosmological history.  The green shaded region for $M_2 < 108 \GeV$ is excluded by the LHC search described above.  
The relevant exclusions from Fermi (red with dashed border) and H.E.S.S. (blue with solid border) are also shown.  Note that the Fermi limit is approximately independent of uncertainties on the profile density because the relevant unknown astrophysical parameters have already been marginalized over, while the H.E.S.S. limit assumes an NFW profile.  A detailed discussion of how the limits depend on the choice of profile is presented in Sec.~\ref{sec: Astro}.

H.E.S.S. places a limit on the total number of line photons from annihilation to $\gamma \,\gamma$ with energy $E_{\gamma\gamma}$.  However, the process $\chi^0\,\chi^0 \rightarrow \gamma \,Z^0$ also produces line photons with energy $E_{\gamma Z}$.  The difference between $E_{\gamma\gamma}$ and $E_{\gamma Z}$ compares to the given resolution of H.E.S.S. as \cite{Abramowski:2013ax}
\begin{equation}
E_{\gamma\gamma} - E_{\gamma Z} = \frac{m_Z^2}{4\,M_2} \quad  < \quad E_\text{res} = 
\left\{
\begin{array}{cc}
0.17\times E_\gamma & \quad\text{for } E_\gamma = 500 \GeV \\
0.11 \times E_\gamma & \text{for } E_\gamma = 10 \TeV
\end{array}\right.
\end{equation}
in the entire probed range of $M_2$.  So, the H.E.S.S. result can be interpreted as a constraint on the sum of the cross section for $\chi^0\,\chi^0 \rightarrow \gamma \,\gamma$ plus half of the cross section for $\chi^0\,\chi^0 \rightarrow \gamma\,Z^0$.    In fact, the $\gamma \,Z^{0}$ final state typically dominates by a factor of 3.  Note that we are neglecting contributions from internal bremsstrahlung, which increase the number of photons contributing to the line signal when energy smearing is taken into account \cite{Bergstrom:2005ss}.

The cross section for $\chi^0\,\chi^0 \rightarrow W^+ W^-$ is plotted as a dashed red line\footnote{The Fermi limit is on the total continuum cross section for a $W^+W^-$ final state, but the shape of the spectrum from annihilations to $Z^0\,Z^0$ is effectively identical.  Because the $Z^0Z^0$ annihilation is loop-level, it is subdominant to $W^+W^-$ and has a negligible effect on the size of the continuum cross section.} (the appropriate scale is shown on the left axis) and the annihilation cross section for $\gamma\,\gamma + \frac{1}{2}\,\gamma\,Z^0$ is plotted as a solid blue line (the appropriate scale is shown on the right axis). Here we calculate cross sections in the ``tree-level-SE'' approximation (see Sec.~\ref{Sec: 1-Loop} for definition).    
Note that one-loop corrections \emph{not} included in the Sommerfeld enhancement have been shown to be surprisingly large; the only study including the full one-loop corrections found a suppression of the cross section involving line photons by a factor of $\sim 3$--4 relative to the tree-level-SE approximation employed here and in the earlier literature \cite{Hryczuk:2011vi}.  We describe the higher-order corrections in detail in Sec.~\ref{Sec: 1-Loop}.

In Fig.~\ref{Fig: Wino Exclusion Thermal}, the yellow region between the dashed black lines corresponds to $\Omega_\text{wino}\,h^2 \in 0.12\pm 0.006$~\cite{Ade:2013zuv}, and is where the wino comprises all of the DM.  We plot a 5\% band, which is dominated by theoretical uncertainty.  Note that the low side of this band is equal to the WMAP 9 year measurement of $\Omega\,h^2$ \cite{Hinshaw:2012aka}.  For masses above the dotted line in the grey region, the wino overcloses the Universe, while for masses below the yellow region, it is a subdominant component of the DM.  A thermal wino with a mass $\sim3.1\TeV$ that accounts for all the dark matter is safely excluded for the NFW profile.\footnote{For reference, if we use the WMAP measurement of the DM relic abundance, a thermal wino has a mass of 3 TeV.}  In contrast, Fig.~\ref{Fig: Wino Exclusion Non Thermal} shows that the non-thermal winos are excluded for the full range of plotted masses by a combination of H.E.S.S., Fermi and the LHC (assuming an NFW profile and no theoretical uncertainty). 

\section{Uncertainties}
\label{sec:uncertainties}
We have shown that Fermi and H.E.S.S. place stringent constraints on the wino parameter space. This section is devoted to exploring four independent issues that could potentially render these limits uncertain, \emph{e.g.} the range of allowed DM density profiles, one-loop corrections to the bare annihilation cross section (which is modified by the SE), temperature-dependent effects in the relic density calculation, and the contribution of higher partial waves. 

\subsection{Dark Matter Density Profile}
\label{sec: Astro}

Astrophysical uncertainties dominate the prediction of the wino annihilation flux.  The flux is proportional to the $J$-factor, defined as 
\begin{eqnarray}
\label{Eq: J}
J &=& \frac{1}{R_{\odot}} \left(\frac{1}{\rho_0}\right)^2 \int_{\Delta \Omega} \mathrm{d}\Omega \int_{\text{los}} \rho^2\big(r\left(s, l, b\right)\big) \,\mathrm{d}s,
\end{eqnarray}
\begin{figure}[b]
\centering
\includegraphics[width=0.45\textwidth]{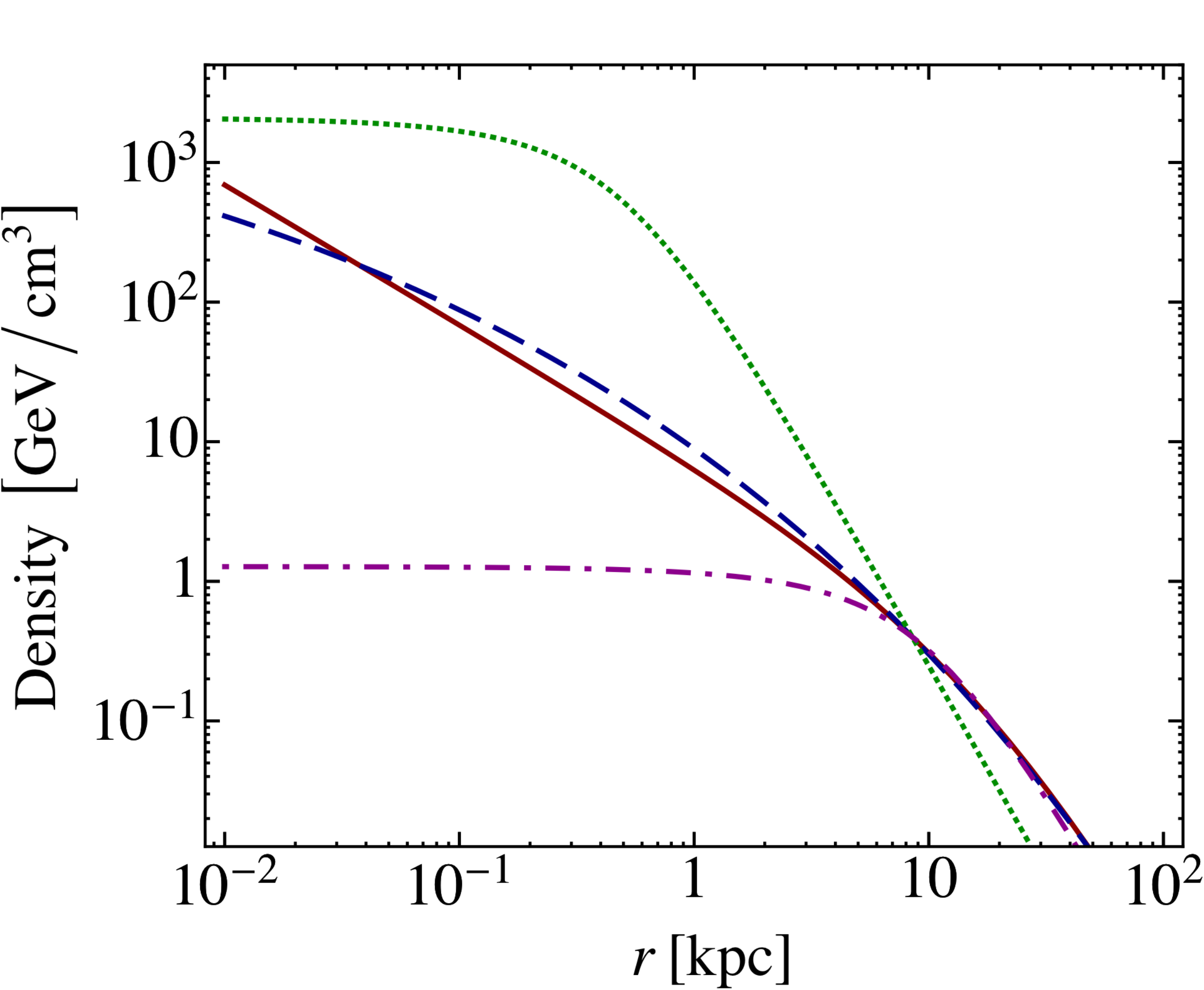}
   \hspace*{.2in}
\raisebox{0.3\height}{\includegraphics[width=0.3\textwidth]{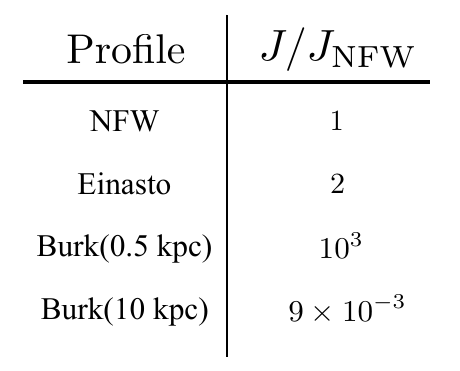}}
\caption{
The NFW [solid, red], Einasto [dashed, blue], and Burkert with $r_s = 0.5$ [green, dotted] and 10 kpc [purple, dot-dashed] profiles as a function of the distance from the Galactic Center.  The table shows the $J$-factors for each of these profiles in the H.E.S.S.~region of interest, normalized to $J_{\text{NFW}} = 0.60$.   
}
\label{Fig: J Factor}
\end{figure}
where $s$ is the line-of-sight distance, $l$ ($b$) is the Galactic longitude (latitude), $r = \sqrt{s^2+R_{\odot}^2-2 s\, R_{\odot}\, \cos l\, \cos b\,}$ is the galactocentric distance, $R_{\odot} = 8.5$ kpc is the distance to the Sun from the Galactic Center, and $\rho_0 = 0.4$ GeV cm$^{-3}$ is the local density~\cite{Catena:2009mf,Weber:2009pt,Salucci:2010qr,Pato:2010yq}.  The functional form for the DM density $\rho(r)$ is highly uncertain.  It is often modeled with the NFW profile~\cite{Navarro:1995iw}
\begin{eqnarray}
\rho_{\text{NFW}}(r) &=& \frac{\rho_0}{\left(r/r_s\right) \left( 1+ r/r_s\right)^2} 
\end{eqnarray}
with $r_s = 20$ kpc.  
Another often cited profile is Einasto \cite{Einasto}, which takes the form 
\begin{eqnarray}
\rho_{\text{Ein}}(r) &=&  \rho_0 \exp\left[-\frac{2}{\gamma} \left( \left( \frac{r}{r_s}\right)^{\gamma} - 1 \right) \right]
\end{eqnarray}
with $r_s = 20$ kpc and $\gamma = 0.17$.  Finally, the Burkert profile \cite{Burkert:1995yz}
\bea
\rho_{\text{Burk}}(r) &=& \frac{\rho_0}{(1+r/r_s)(1+(r/r_s)^2)}
\eea
is an example of a cored profile that results in a large range of predictions for the $J$-factor for different choices of $r_s$. The NFW and Einasto profiles are favored by $N$-body dark matter only simulations,\footnote{These $N$-body simulations only include collisionless dark matter.  Recent work suggests that baryonic processes can substantially modify the inner structure of dark matter halos, either flattening or steepening them.  Milky-Way-like halos in simulations that model these processes have been found to possess NFW-like profiles into $\sim 2$ kpc from the GC \cite{DiCintio:2013qxa}, although a larger $\sim 10$ kpc core has been found in one simulation \cite{Maccio:2011eh}.} see for example~\cite{Pieri:2009je}, but there is observational evidence for shallower or cored profiles in some dwarf galaxies~\cite{Walker:2011zu}.

\begin{figure}[tb]
\centering
\includegraphics[width=0.55\textwidth]{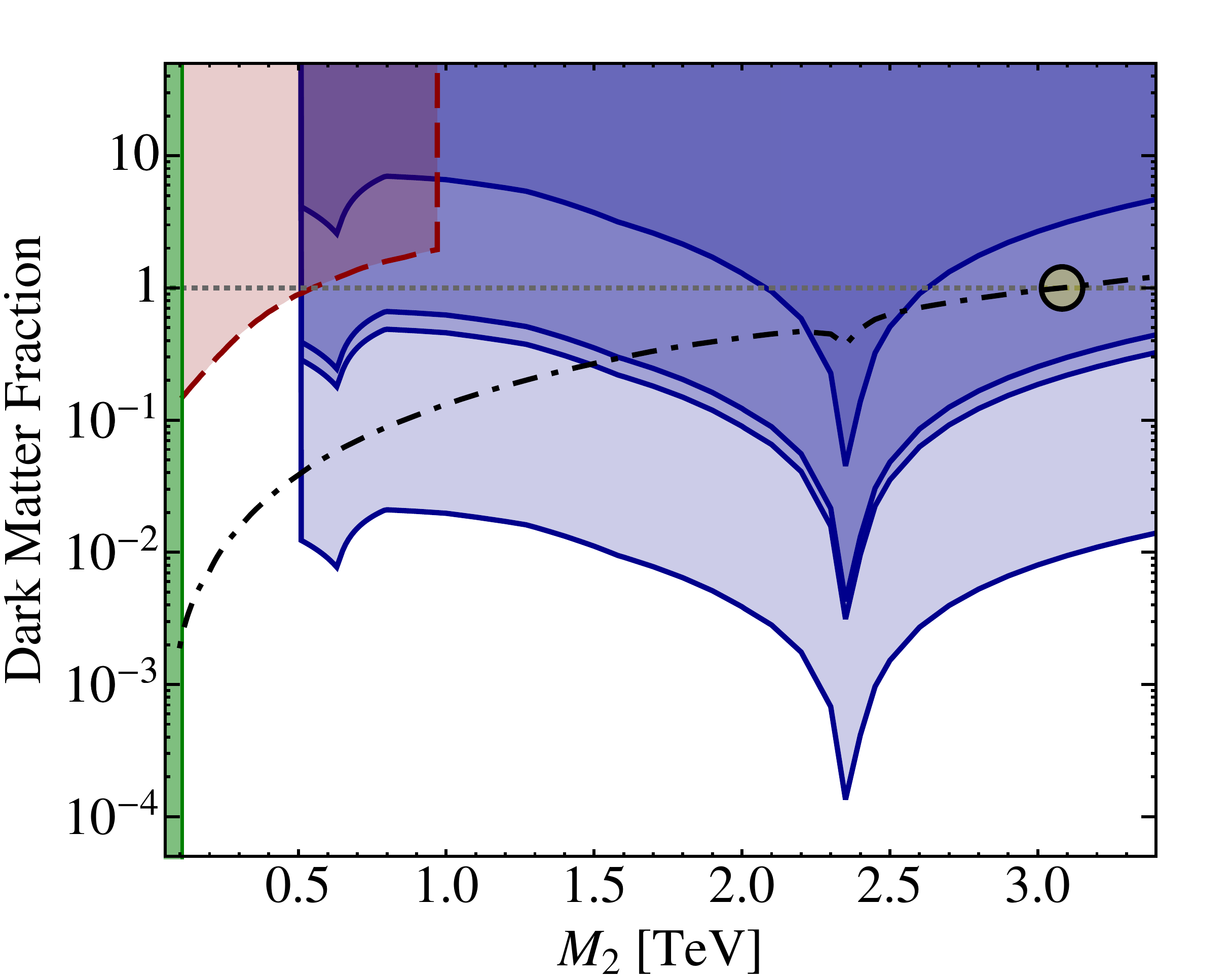}
\caption{
The current bounds from H.E.S.S. [blue, solid] and Fermi [red, dashed] for Burk(0.5 kpc), Einasto, NFW, and Burk(10 kpc) [bottom to top].  The green band is excluded by direct searches at the LHC and the yellow shaded circle corresponds to the thermal wino scenario.  The dotted grey line demarcates where the DM fraction constitutes all of the relic density.  The dot-dashed black line represents the fraction of the DM predicted by a thermal cosmological history.  All cross sections are computed in the tree-level-SE approximation.  \emph{One-loop effects have been shown to reduce the cross section to line photons by as much as a factor of 4} (see Sec.~\ref{Sec: 1-Loop}).
}
\label{Fig: Fraction Excluded}
\end{figure}

These different density profiles are illustrated in Fig.~\ref{Fig: J Factor} and the table lists the corresponding $J$-factors in the H.E.S.S.~region of interest, which is a $1^{\circ}$ circle at the Galactic Center, with the Galactic plane masked out ($|b| \geq 0.3^{\circ}$).   The $J$-factor can vary over several orders of magnitude, depending on the profile.  Microlensing and dynamical observations of the Galaxy appear to be consistent with NFW and Einasto profiles, assuming $r_s = 20$ kpc~\cite{Iocco:2011jz}.  A different analysis using only kinematic observations constrains $r_s = 18 \pm 4.3$ kpc~\cite{McMillan:2011wd} for an NFW profile.  A rough constraint on the core size of a Burkert profile can be obtained from~\cite{Deason:2012wm}.  This study uses Blue Horizontal Branch stars at Galactocentric distances of 16 to 48 kpc as kinematic tracers and places a constraint on the power-law index for the density profile.  The result is consistent with a maximum core size of $\sim 10$ kpc for the Burkert profile, though this requires a significant extrapolation of 
the analysis down to much lower distances.  Thus, the $J$-factors listed in Fig.~\ref{Fig: J Factor} correspond to profiles that are all consistent with current constraints.  Until dynamical measurements can determine the DM profile of the Milky Way to greater precision, $J$-factor uncertainties over several orders of magnitude will remain.

Fig.~\ref{Fig: Fraction Excluded} shows the DM fraction excluded by H.E.S.S. for Burk(0.5 kpc), Einasto, NFW, and Burk(10 kpc) profiles.  The Fermi exclusion, which is marginalized over profile uncertainty, is also plotted.  Near the resonance at  2.3 TeV, a  wino is excluded independent of profile if it is the dominant DM.
A thermal wino that makes up all of the DM is excluded for the Burk(0.5 kpc), Einasto, and NFW profiles at present (see the yellow shaded circle of Fig.~\ref{Fig: Fraction Excluded}), but not yet for a profile with a large core, \emph{e.g.} Burk(10 kpc). 

\subsection{Loop Corrections to Annihilation}
\label{Sec: 1-Loop}
The Sommerfeld enhancement for neutral winos can be expressed as a multiplicative matrix factor, which is to be contracted with the matrix describing the ``bare'' annihilation rate for charginos and neutralinos; Appendix~\ref{Appendix: Sommerfelding} contains a detailed review of the Sommerfeld calculation. However, this formalism does not account for all the one-loop contributions to the annihilation cross sections. It is important to investigate the impact of these one-loop perturbative corrections on the full non-perturbative cross section~\cite{Hryczuk:2011vi}, especially for the production of line signals where the annihilation cross section is zero at tree-level.  For concreteness, we discuss annihilations to $\gamma \gamma$ in this section -- the same story holds for the $\gamma\,Z^0$ final state that is responsible for the dominant line signal.  

Including the leading contributions, the one-loop-SE cross-section for wino annihilation to $\gamma \gamma$ is (as derived in Appendix~\ref{Appendix: Sommerfelding}):
\begin{align} \label{Eq: Sigma^GammaGamma}
\sigma_{00}\, v
& = 2 \bigintssss \Bigg(\Big|s_{00,+-}\Big|^2 \left( \Big |\mathcal{A}^{\gamma \gamma}_{+-}\big(g^2\big) \Big |^2 + 2 \,\mathrm{Re} \Big[\mathcal{A}^{\gamma \gamma\,*}_{+-}\big(g^2\big)\,\mathcal{A}^{\gamma\gamma}_{+-}\big(g^4\big)\Big] \right)  \nonumber \\ 
&\quad\quad\quad\quad\quad\quad\quad + \,2 \,\mathrm{Re} \Big[s_{00,00}\,s_{00,+-}^*\mathcal{A}^{\gamma \gamma\,*}_{+-}\big(g^2\big)\,\mathcal{A}^{\gamma\gamma}_{00}\big(g^4\big)\Big] \Bigg) + \OO\big(g^8\big).
\end{align}
Here, $\sigma_{00}$ is the neutralino-neutralino annihilation cross section, $v$ is the relative velocity of the annihilating particles, $\mathcal{A}^{\gamma\gamma}_j(g^n)$ denotes the hard matrix element to $n^{\text{th}}$ order in the gauge coupling $g$ for the annihilation of the initial state $j$; $s_{ij}$ is the SE associated with the two-body state $i$ becoming state $j$, and the integral is taken over phase space.  We refer to the cross section computed using \eref{Eq: Sigma^GammaGamma} as ``one-loop-SE," which contrasts with the $\OO\big(g^4\big)$ plus SE calculation of \cite{Hisano:2006nn} that is used to compute the cross sections presented in all the above figures (referred to as ``tree-level-SE"), and the $\OO\big(g^8\big)$ perturbative calculation without any SE contribution (referred to as ``one-loop-perturbative"). For the neutral wino, the tree-level-SE calculation is equivalent to only retaining the first term in \eref{Eq: Sigma^GammaGamma}, as discussed in Appendix~\ref{Appendix: Subtraction}.

Note that the unenhanced one-loop-perturbative cross section \cite{Bergstrom:1997fh, Bern:1997ng} is $\OO\big(g^8\big)$ and is not directly included in \eref{Eq: Sigma^GammaGamma}. However, the inclusion of the SE numerically captures the leading portion of this contribution at large $M_{2}$.  At these large masses, the residual $\OO\big(g^8\big)$ piece of the perturbative cross section is subdominant.

There is a subtlety that must be accounted for when computing the higher-order terms in Eq. (\ref{Eq: Sigma^GammaGamma}).  Specifically, the non-relativistic limit of the one-loop amplitude involving a single ladder-diagram-like $W^\pm$ exchange should be subtracted from the full one-loop amplitude before including it in the annihilation matrix because this diagram is already included in the SE factor.  The explicit subtraction procedure used by \cite{Hryczuk:2011vi} is reviewed in Appendix \ref{Appendix: Subtraction}.  

The effect of this subtraction is to completely remove the part of the $\mathcal{A}^{\gamma\gamma}_{00}$ amplitude proportional to $M_2/m_W$, 
which at high masses would give rise to the leading contribution to the one-loop cross section for neutralino annihilation to $\gamma \gamma$. In other words, the usual $\alpha^2 \,\alpha_W^2/m_W^2$ scaling of the one-loop $\gamma \gamma$ line cross section \cite{Bern:1997ng} is due entirely to the one-loop SE.  Once this subtraction is performed, the residual amplitude (with the tree-level amplitude removed as in \cite{Hryczuk:2011vi}) is a function of $\log\left(M_2/m_W\right)$ and $\log^2\left(M_2/m_W\right)$ \cite{Hryczuk:2011vi, Hryczuk}.  

The analogous subtraction must also be performed for the one-loop amplitude for chargino annihilation into photons, $A^{\gamma \gamma}_{+-}(g^4)$.  Again the residual amplitude is only a function of $\log\left(M_2/m_W\right)$ and $\log^2\left(M_2/m_W\right)$ \cite{Hryczuk:2011vi, Hryczuk}.  However, as shown in Fig.~11 of \cite{Hryczuk:2011vi}, this residual one-loop amplitude can still be very large as a fraction of the tree-level amplitude, approaching $40\%$ for 3 TeV DM -- this is attributed to accidentally large $\OO(10)$ coefficients that multiply $\log^2\left(M_2/m_W\right) \sim 10$ and compensate for the loop factor suppression\footnote{As is noted in \cite{Hryczuk:2011vi}, for $M_2 \gg 3 \TeV$,  perturbation theory is breaking down and in order to extrapolate this calculation to higher masses these large logs must be resummed.} \cite{Hryczuk}.  This leads to a suppression of the line signal, at the low velocities relevant to the Milky Way halo, by a factor of $\sim 3$--4.  Because going to the next order in perturbation theory leads to an $\OO(1)$ change in the cross section due to the presence of large logs, it may be important to work in a non-relativistic effective theory that would allow resummation of these effects.  This will be investigated in future work.

\begin{figure}[t!] 
   \centering
  \includegraphics[width=0.55\textwidth]{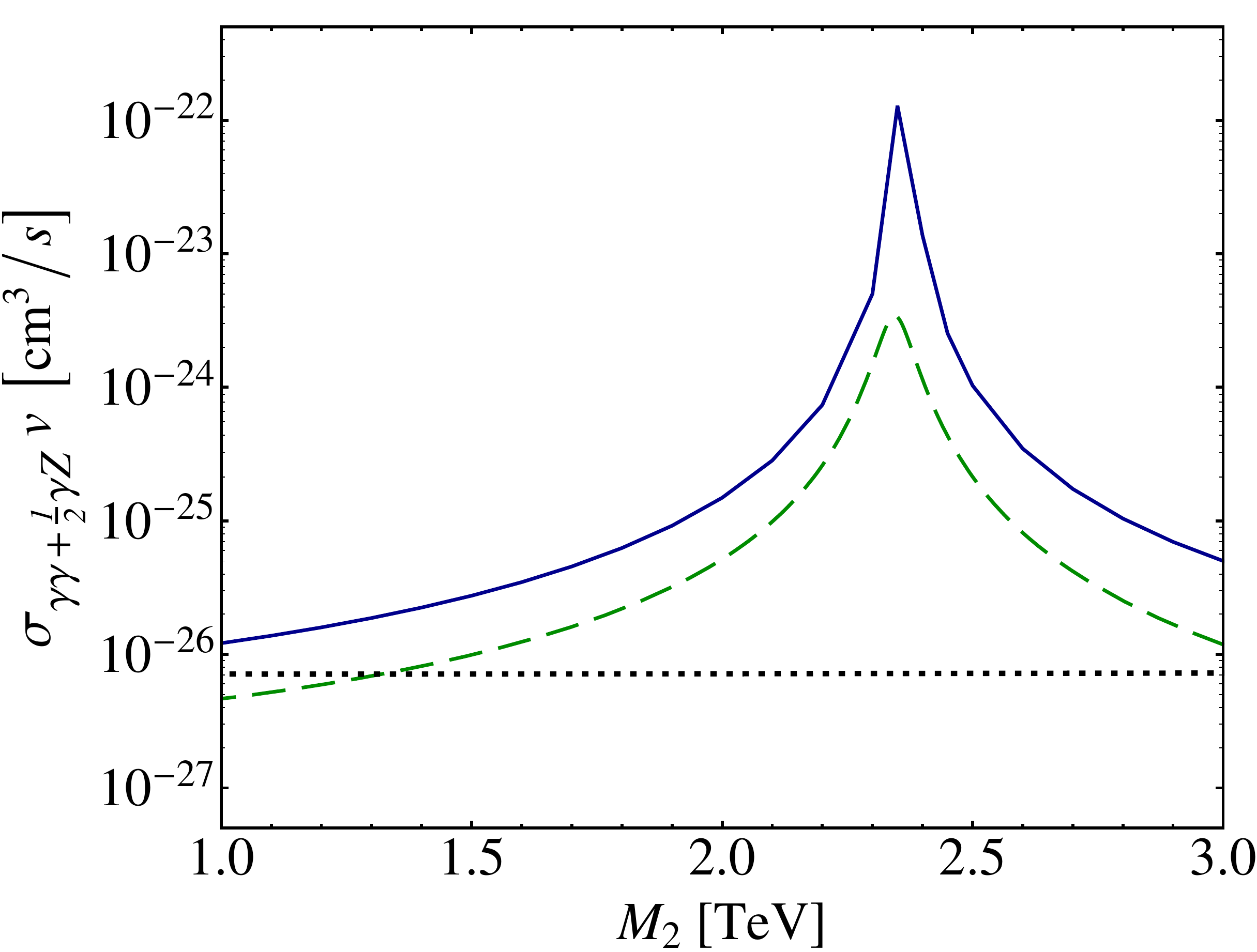} 
     \caption{The neutral wino annihilation to line photons as computed with three approximations: one-loop-perturbative [black, dotted]; tree-level-SE [blue, solid]; one-loop-SE [green, dashed].}
   \label{Fig: Compare GG Xsec}
\end{figure}

Figure~\ref{Fig: Compare GG Xsec}  compares the cross section for neutralino annihilation to line photons for three different approximations: one-loop-SE (green, dashed) \cite{Hryczuk:2011vi} (\emph{i.e.}, \eref{Eq: Sigma^GammaGamma}), tree-level-SE (blue, solid) \cite{Hisano:2004ds}, and one-loop-perturbative (black, dotted) \cite{Bergstrom:1997fh}.  Note that we calculate the tree-level-SE result ourselves while the one-loop-perturbative result is computed using \texttt{DarkSUSY} \cite{Gondolo:2004sc}, and the the one-loop-SE curve is taken from~\cite{Hryczuk:2011vi}.  

The annihilation cross section to $W^+W^-$ has also been computed to $\OO\big(g^6\big)$ \cite{Hryczuk:2011vi}, and the impact of the higher-order correction is smaller than that for photon annihilation, ranging from $\sim10$--30\%, depending on the wino mass.  The annihilation rate is always dominated by the tree-level cross section enhanced by the $|s_{00,00}|^2$ Sommerfeld factor -- although the large $\log (M_2/m_W)$ factors do still contribute to the 1-loop result, and are the reason this correction is as large as observed.  Note that our limit plots use the tree-level-SE calculation for $W^+W^-$ annihilation and thus do not include this small uncertainty.

The neutral wino annihilation to $W^+W^-$, along with the tree-level contributions to $\chi^+ \chi^-$ annihilation, is the main contribution to the total annihilation cross section above the threshold for on-shell production of $\chi^+ \chi^-$ and hence also controls the relic density. For calibration, at low velocities, the full $\chi^0\, \chi^0 \rightarrow W^+ \,W^-$ cross section is greater than  $\chi^0\, \chi^0 \rightarrow Z^0\, Z^0$ by about an order of magnitude. Thus,  the effect of one-loop corrections on the relic density should be modest, around $\sim 20$--30\%.  Also, because the one-loop corrections reduce the total annihilation cross section, these effects reduce the mass that gives a thermal wino with the correct relic density.  Figure~\ref{Fig: Wino Exclusion Thermal} shows that this corresponds to a tighter bound on the thermal wino, as lower masses are closer to the resonance region.

\subsection{Temperature-Dependent Effects}
For wino masses in the TeV-range,  freeze-out occurs at $x_f \simeq 20$, giving $T_f \sim 100 \GeV$.  For temperatures of this order, the Higgs vacuum expectation value (vev) can still be adiabatically transitioning from the electroweak-preserving vacuum to its zero temperature value \cite{D'Onofrio:2012ni}.  Therefore, the dynamics of the electroweak phase transition can potentially affect the physics of freeze-out \cite{Cirelli:2007xd}.  In this section, we will argue that uncertainty introduced by ignoring the temperature dependence of the masses and couplings is small.

These effects can manifest in a variety of ways.  Because the mass splitting between the charged and neutral wino is proportional to $m_W$, the mass splitting $\delta$ vanishes at high temperatures when electroweak symmetry is restored.  The gauge boson masses and interactions are also affected by the presence of finite temperature.  The transverse polarizations of the gauge bosons have a mass set by the Higgs vev and these modes become massless at high temperature.  The timelike polarizations of the gauge bosons receive Debye masses proportional to $T^2$.  This can be understood from the picture of finite temperature as a compactification of the time direction in the four dimensional theory onto a circle.  The timelike modes are scalars in the theory on $\mathbb{R}^3\times S^1$ and can consistently obtain temperature-dependent masses.  For a given temperature, the $B_0$--$W^3_0$ matrix must be diagonalized and the interactions between the winos and the timelike modes become temperature-dependent.  
Previous calculations included a large portion of these effects by modifying the gauge boson masses \cite{Cirelli:2007xd}.

\begin{figure}[tb]
\centering
\includegraphics[width=0.48\textwidth]{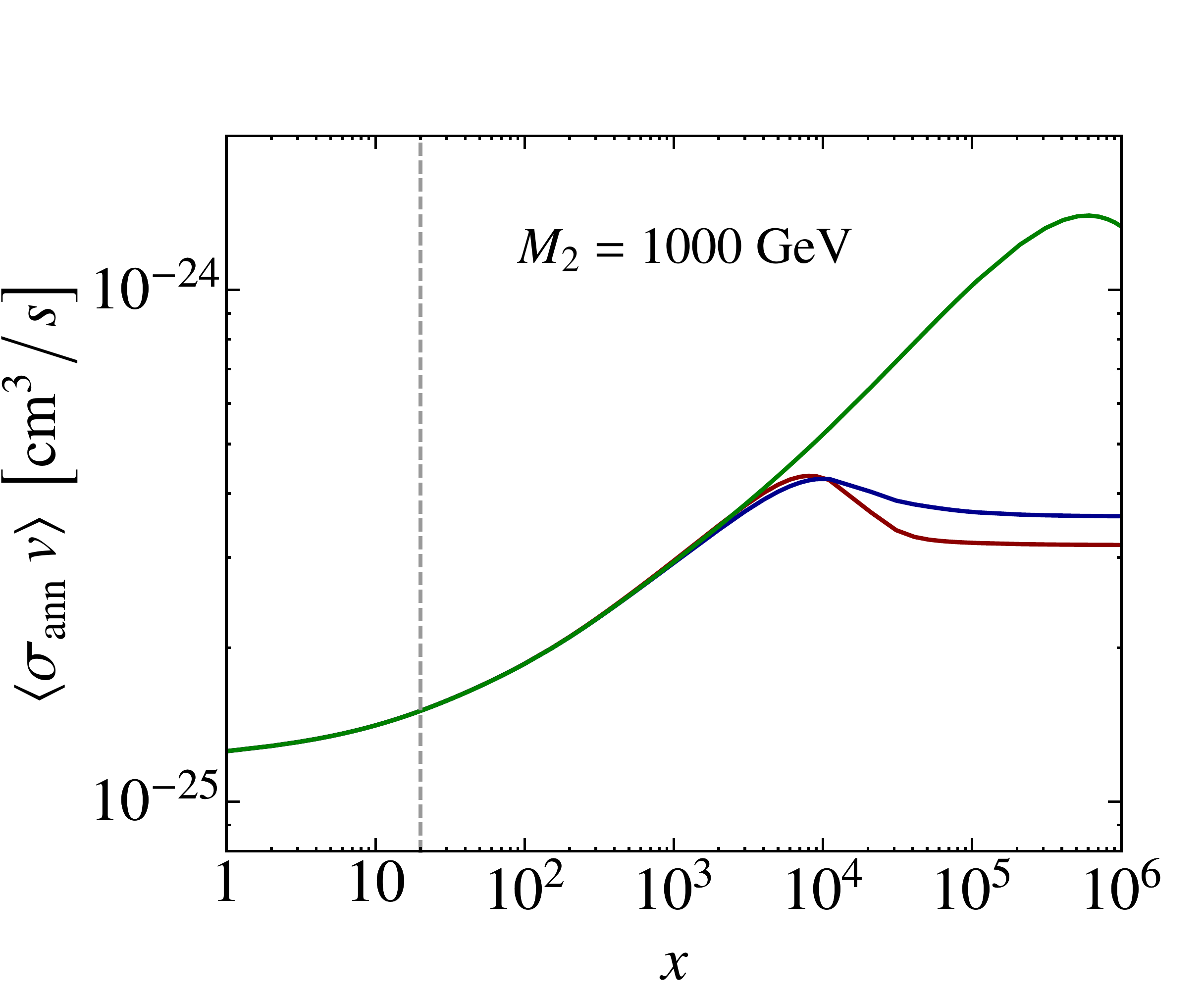} $\!$ \includegraphics[width=0.48\textwidth]{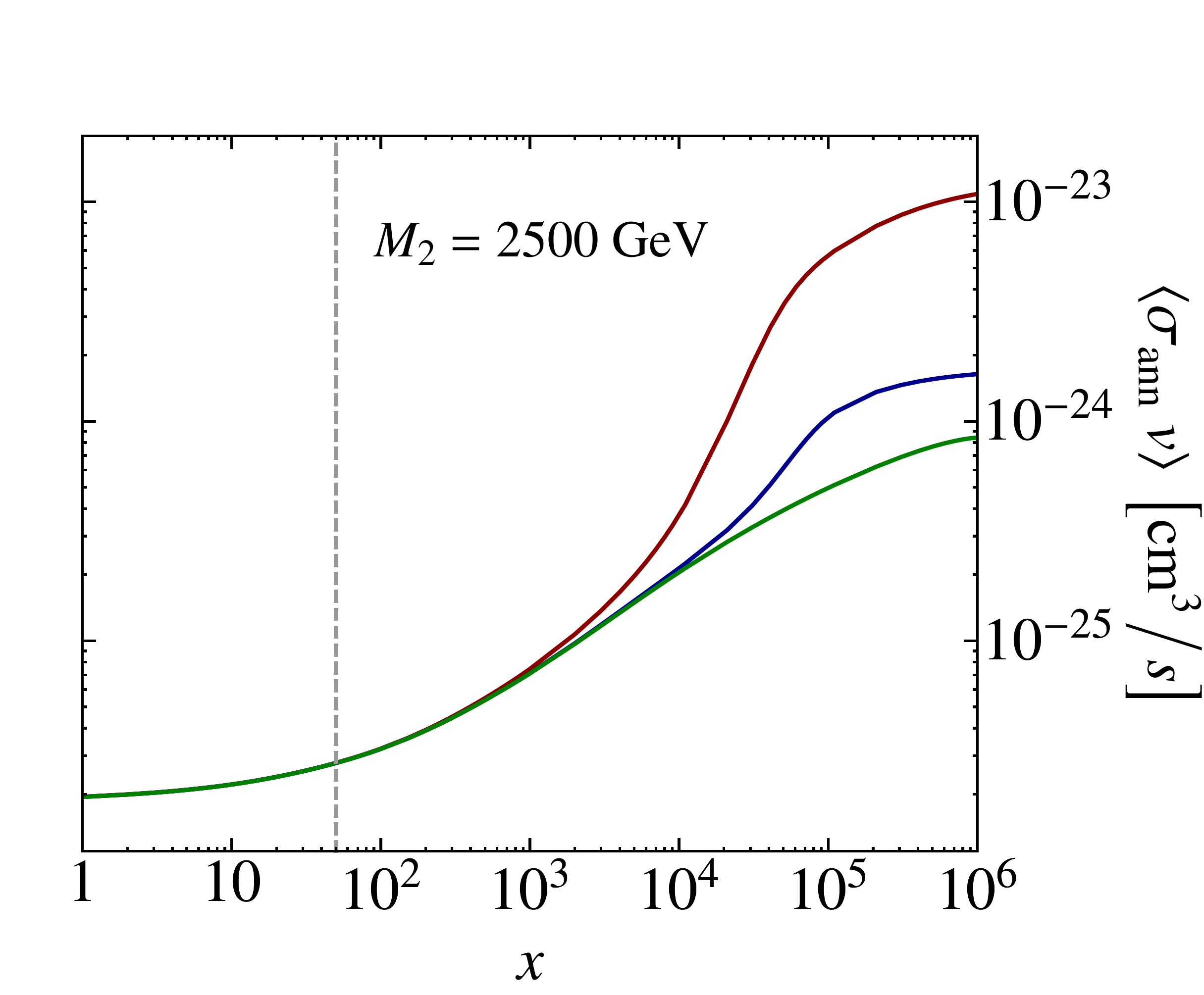}
\caption{
The thermally averaged tree-level-SE cross section as a function of $x \equiv M_2/T$.  On the \emph{left}, the wino mass is 1 TeV and the solid lines (from top to bottom) show $\delta = 1 \MeV, 0.1 \GeV, \text{ and } 0.17 \GeV$.  On the \emph{right}, the wino mass is 2.5 TeV and the solid lines show $\delta = 0.17 \GeV, 0.1 \GeV, \text{ and } 1 \MeV$ from top to bottom, respectively.  The dotted vertical line demarcates $x_\text{PT}$ assuming $T_\text{PT} = 50 \GeV$.
}
\label{Fig: TDependence}
\end{figure}

No full calculation exists to show the impact of these temperature-dependent effects on the relic density calculation.  Therefore, we performed a variety of tests to determine the maximum impact that could result from the phase transition, 
which is approximated as a sharp change in the parameters of the model at $T_{\text{PT}} = 50 \GeV$.  Figure~\ref{Fig: TDependence} illustrates the effect on the thermally averaged tree-level-SE cross section as the mass splitting $\delta$ is varied at ``high'' temperature, \emph{i.e.}, before the mock phase transition completes at 50 GeV.  The left (right) panel shows the result for $M_2 = 1 \TeV\,(2.5 \TeV)$.  The $x$ value that corresponds to $T_{\text{PT}} = 50 \GeV$ is demarcated by the vertical grey dotted line.  For small $x$ (high temperatures), these curves are indistinguishable -- any temperature dependence in $\delta$ has no effect on the relic density.

The temperature dependence on the gauge boson masses and couplings may also be relevant.  As it turns out, there is no sizable impact on the relic density if the  electroweak gauge boson masses $m_W$ and $m_Z$ are reduced above $T_{\text{PT}}=50$ GeV.  This makes sense because, at these high temperatures, the gauge boson masses can be neglected.  Similarly, there is no effect on the relic density when the values of $\alpha$ and $\alpha_W$ are reduced.  This reduction increases the DM abundance before the phase transition, but once the temperature drops below 50 GeV, the full-strength annihilations deplete the DM density back to the usual thermal relic value.  Note that if any temperature-dependent effects lead to a \emph{depletion} of the DM before the phase transition, there could be a sizable change in the relic density.  However, this is not the case for the effects described above and we conclude that temperature-dependent effects are subdominant to one-loop effects and profile uncertainties.

\subsection{Velocity Suppressed Contributions}

In both the relic density and present-day annihilation calculations, $\mathcal{O}\big(v^2\big)$ and higher contributions to the perturbative annihilation cross section have been ignored, \emph{i.e.}, both $p$-wave terms and subdominant $s$-wave terms have been neglected.\footnote{A formalism for separating $\mathcal{O}\big(v^2\big)$ $s$-wave from $p$-wave has been developed in \cite{Hellmann:2013jxa}.} The effect of velocity-suppressed terms on the bare annihilation cross section is completely negligible, as typical halo velocities are $\sim 10^{-3} \,c$.  Because freezeout occurs at $M_2/T \sim 20$, the impact of these terms on the relic density is $\mathcal{O}\big(10$--$15\%\big)$.  These effects modify the present-day signal for winos with a thermal history by  $\mathcal{O}\big(20$--$30\%\big)$, and also increase the mass at which the thermal wino constitutes $100\%$ of the DM by $\sim0.2$ TeV.  Because this effect is rather small compared to the other uncertainties discussed above, we feel comfortable neglecting these corrections.

One might ask whether the velocity-dependent SE changes these parametric statements. To examine the effect of \emph{non}-resonant SE, which applies an $\mathcal{O}(\alpha/v)$ enhancement to the annihilation rate in the $s$-wave case, we can use the results for the Coulomb potential, which behaves similarly to the true potential at higher velocities where the gauge boson masses and mass splitting can be largely neglected. For a Coulomb potential with coupling $\alpha$, the SE for partial waves $\ell$ is given by \cite{Cassel:2009wt}
\begin{equation} 
S_{\ell > 0} = \left( \frac{2\,\pi \,\alpha}{v} \frac{1}{1 - e^{-2 \pi \alpha/v}} \right) \times \prod_{n=1}^\ell \left(1 + \frac{(\alpha/v)^2}{\,n^2} \right).
\end{equation}
The effect of the non-resonant SE is to cancel out the $\sigma\, v \sim v^{2\, \ell}$ dependence, for $v \ll \alpha$.  Thus, for all partial waves, the cross section $\sigma \,v$ scales as $1/v$ for the Coulomb potential at sufficiently low velocities.
However, the $v^{2\,\ell}$ suppression of the $\ell$-wave is effectively \emph{replaced} by an $\alpha^{2\,\ell}$ suppression for $v \ll \alpha$. This implies that for small $\alpha$ and small $v$, the higher partial-wave terms are still subdominant to the $s$-wave.

For finite-range potentials, the enhancement will saturate when $M_2\, v$ becomes comparable to the force carrier mass; below this velocity, the usual $v^{2 \,\ell}$ dependence of $\sigma_\ell \,v$ will be recovered (although the value of the enhancement at saturation will scale as $(\alpha / v_\mathrm{saturation})^{2 \,\ell}$, as discussed above). In no case can the non-resonant Sommerfeld enhancement cause the higher partial waves to be un-suppressed at low velocities. We have confirmed by direct numerical calculation that, for the parameter space of greatest interest with $M_2 < 10$ TeV, the $p$-wave enhancement is always comparable to the $s$-wave enhancement or smaller.

Another concern is whether the higher-order velocity contributions experience a different SE resonance structure.  
Resonances occur when the potential develops a bound state at zero energy \cite{Hisano:2004ds}, enhancing the annihilation of particles in near-zero-energy, \emph{i.e.}, low velocity, scattering states.   At sufficiently low DM masses (below the first resonance), the potential has no bound states at all; the first resonance corresponds to the appearance of the first bound state in the spectrum. As the mass of the DM is increased, holding the other parameters fixed, more bound states develop,  provided the potential is attractive.  Each new bound state causes a resonance when it appears  because its energy is very close to zero when it first becomes bound.

The higher partial waves may have bound states that are not degenerate with the $s$-wave bound states, leading to $p$-wave (or higher) resonances that would appear at a different mass than in the $s$-wave calculation. In this case, one might worry that higher partial waves could have a dramatic effect on the results for the relic density and/or the present-day signal.
However, the bound states for the higher partial waves are always more shallowly bound than for the $s$-wave.

The single-state Yukawa potential has bound states for higher partial waves that are nearly degenerate with the $s$-wave bound states, but there is no ``leading'' $p$-wave bound state corresponding to the lowest $s$-wave bound state. Consequently, there is no $p$-wave resonance in the same region of DM mass as the first $s$-wave resonance (the first $p$-wave resonance is close to the second $s$-wave resonance, roughly a factor of 4 higher in mass). This can be seen both by comparing Fig.~3 and Fig.~4 of \cite{Cassel:2009wt}, and by studying the analytic approximation in that work. 

The absence of a leading $p$-wave bound state can be generalized to the more complicated multi-state potential for neutral winos.  Following the notation of Appendix \ref{Appendix: Sommerfelding}, for the $(Q=0, S=0)$ system, the potential for the wino is bounded below by the related potential,
\begin{equation} 
V(r) =  -\sqrt{2} \,\frac{\alpha_W\, e^{-m_W\, r}}{r}\, \left( \begin{matrix} 1 & 1 \\ 1 & 1 \end{matrix} \right).
\end{equation} 
Note that there can only be a bound state in the wino system if one exists for this deeper potential (equivalently, the ground state energy of this potential is lower than for the wino). However, the Schr{\"o}dinger equation for this potential can be diagonalized, yielding two uncoupled equations for the eigenstates; one experiences no potential ($V=0$), while the other experiences 
\be
V(r) = - 2\, \sqrt{2}\, \alpha_W\, \frac{e^{-m_W \,r}}{r}.
\ee
The first $p$-wave bound state for the latter potential appears when \cite{Cassel:2009wt},
\be
m_\chi \sim \frac{7 \, m_W}{2\, \sqrt{2}\, \alpha_W} \sim 6  \TeV.
\ee   Hence, there will be no $p$-wave (or higher $\ell$) bound states for neutral wino masses below $\sim 6$ TeV -- numerical computation indicates that the first $p$-wave resonance occurs at $m_\chi \sim 11$ TeV.
This justifies neglecting the higher partial waves in this work.

\section{Future Projections}
\label{sec:projections}

The Cherenkov Telescope Array (CTA) experiment \cite{Consortium:2010bc} is a next-generation ground-based gamma-ray observatory, with data expected in 2018.  Its design represents a dramatic increase in effective area over the H.E.S.S. experiment and it consequently has a much improved reach in the gamma-line search.  Here, we present a projection of its capabilities \cite{Bergstrom:2012vd}.  The projection is based on a log-likelihood analysis, and relies on a  relatively modest 5 hours of observing time, with an (energy-dependent) effective area given as in \cite{Vandenbroucke:2011sh}, and energy resolution given as in \cite{Consortium:2010bc}.     The Galactic Center background is parameterized as in \cite{Bringmann:2011ye}, and is an admixture of mis-identified protons, diffuse gamma-rays, and cosmic-ray electrons.  The observation time is chosen to keep the limits in the statistically-limited regime: the limits presented approach signal to background of 1\%.   

\begin{figure}[tb]
\centering
\includegraphics[width=0.48\textwidth]{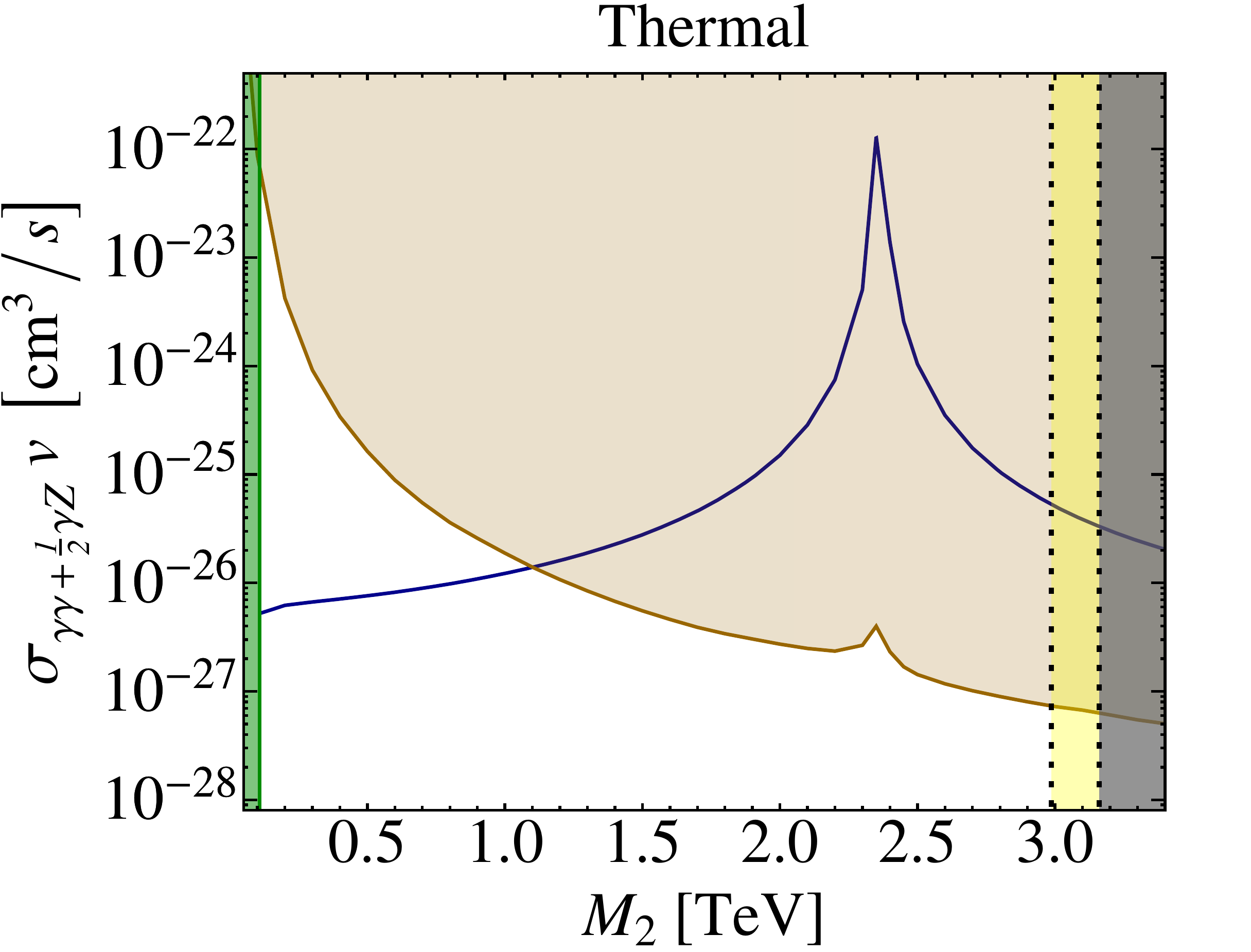} $\!$ \includegraphics[width=0.48\textwidth]{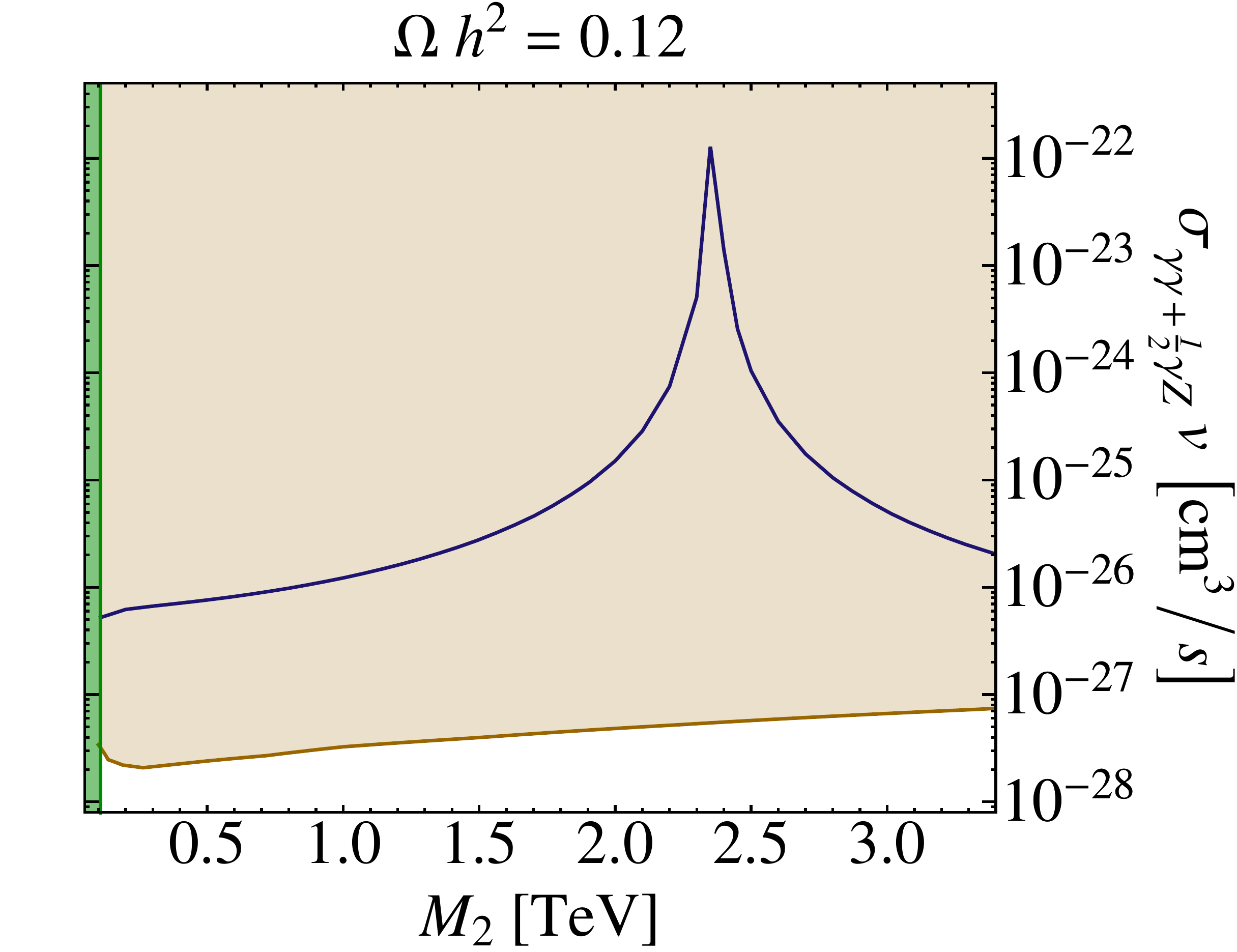}
\caption{
The same as Figs.~\ref{Fig: Wino Exclusion Thermal} and \ref{Fig: Wino Exclusion Non Thermal} except that  the orange shaded regions are for the 5 hour CTA projection of \cite{Bergstrom:2012vd, Weniger}.
}
\label{Fig: CTA projections}
\end{figure}

\begin{figure}[t!]
\centering
\includegraphics[width=0.58\textwidth]{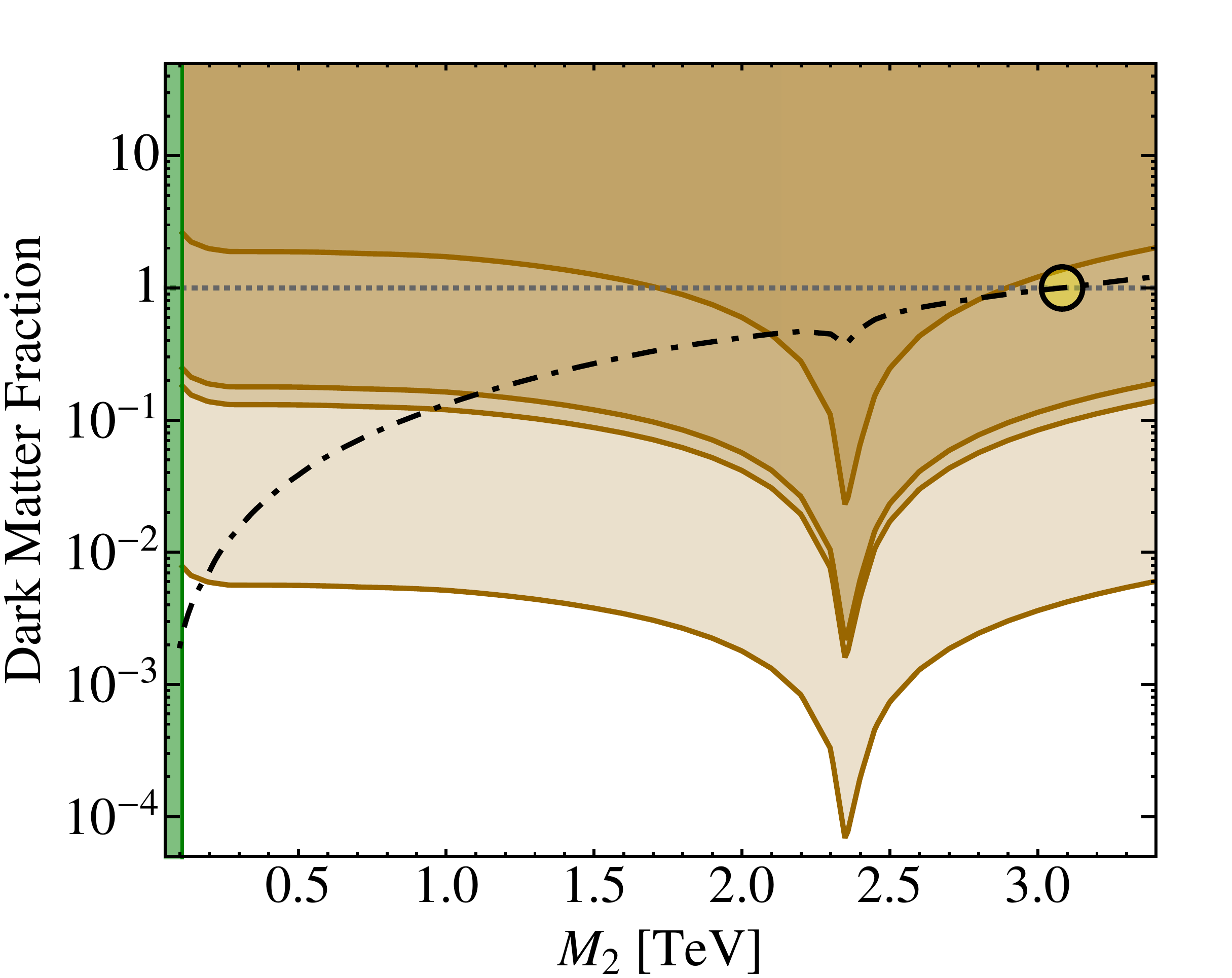}
\caption{The same as Fig.~\ref{Fig: Fraction Excluded}, except that the orange shaded regions are for the 5 hour CTA projection of \cite{Bergstrom:2012vd, Weniger}.
}
\label{Fig: Fraction Excluded CTA}
\end{figure}

As Fig.~\ref{Fig: CTA projections} shows, the projected limits are powerful.  For a thermal wino that provides the full relic density of DM ($M_2 \simeq 3.1$ TeV), CTA will exclude the tree-level-SE cross section by a factor of $\sim 60$ for an NFW profile.   Indeed, from examining the left panel of Fig.~\ref{Fig: CTA projections}, a wino with a thermal abundance is excluded all the way down to 1.1 TeV, where it makes up only $\sim  16\%$ of the total relic abundance.  The right panel shows that a wino making up the full relic density of the DM  -- independent of the cosmological history -- would be robustly excluded over the entire mass range shown.  As Fig.~\ref{Fig: Fraction Excluded CTA} shows, only the most pessimistic DM profiles would evade detection.  

Measurements of the anti-proton flux from AMS-02 will continue to tighten the constraints on wino annihilations to $W^+W^-$ beyond those obtained from Fermi gamma-ray and PAMELA anti-proton measurements.  The estimated reach of AMS-02 after 1 and 3 years of data is given in~\cite{Cirelli:2013hv}.  On their own, the AMS-02 results should exclude winos that comprise all the DM from $100 \GeV \lesssim M_2 \lesssim 3.1 \TeV$, and thermal winos with mass below 200 GeV (for the NFW profile and in the tree-plus-SE approximation).  The anti-proton propagation is an additional source of systematic error for AMS-02, resulting in an uncertainty of one to two orders of magnitude in the limit estimation.

\section{Conclusions}
\label{sec:conclusions}

In this paper, we explored the limits on wino DM.  Thermal winos comprise all of the DM at a mass of $\sim3.1\TeV$; this provides a motivation for the presence of gauginos at the weak scale in models with split supersymmetry spectra.  Although collider and direct detection prospects for TeV-scale wino DM are limited, we have shown that Cherenkov telescopes such as H.E.S.S. and (in the future) CTA are remarkably powerful at exploring this well-motivated DM candidate.

Assuming a thermal history, winos are excluded by H.E.S.S. from 3.1 TeV, where they comprise all of the DM, down to $\sim1.6$ TeV for an NFW profile.  Assuming a non-trivial cosmology, where some additional process is required to keep the wino density at $\Omega\, h^2 = 0.12$ for a given mass, H.E.S.S. excludes winos down to $500\GeV$ for an NFW profile; the Fermi constraint on continuum annihilation to $W^+W^-$ from observations of dwarf spheroidals excludes masses below $500\GeV$.      

These limits are highly sensitive to uncertainties in the DM density profile.  For example, the line photon annihilation cross section for a 3.1 TeV wino is excluded to 95\% confidence by factors of $\sim$12, 22, and 12000 for NFW, Einasto, and Burk(0.5 kpc) profiles, respectively.  It is not excluded for a Burkert profile with 10 kpc core by more than an order of magnitude.  However, winos near the Sommerfeld resonance at $\sim 2.4 \TeV$ are safely excluded for these four profiles.

All cross sections were computed in the tree-plus-SE approximation. The effect of 1-loop corrections not included in the SE calculation may weaken the constraints by up to a factor of 3--4; however, even in this case, a wino constituting the full DM relic density is ruled out over the entire mass range for the NFW profile, and the limit on thermal winos is only raised from 1.6 to 1.7 TeV.

CTA will push these limits even further down, constraining the entire mass region explored here for non-thermal wino production, and excluding winos with mass above $\sim1.1 \TeV$ for thermal histories.  It has the potential to exclude the line photon annihilation cross section for a 3.1 TeV wino to 95\% confidence by factors of $\sim$
60, 110, and 50000 for an NFW, Einasto, and Burk(0.5 kpc) profiles, respectively.  The Burkert profile with a 10 kpc core remains out of reach.  These projections assume only 5 hours of observing time.  

Cherenkov telescopes clearly play an important role in the search for TeV-scale DM.  In the case of wino DM, where collider and direct detection searches have negligible sensitivity, gamma-ray telescopes such as H.E.S.S., Fermi, and CTA provide the most promising window for detection.

\section*{Acknowledgements}
We thank N. Arkani-Hamed, J. Bovy, J. Fan, J. Hisano, A. Hryczuk, R. Iengo, J. Mardon, Y. Nomura, M. Peskin, M. Reece, A. Strumia, J. Wacker, N. Weiner and H. Yu for helpful discussions.  We are especially grateful to C. Weniger for providing us with the projected CTA limits, as calculated in~\cite{Bergstrom:2012vd}, and extending them up to $5\TeV$ for us \cite{Weniger}, and to  A.~Hryczuk and R.~Iengo for extensive discussions of Ref.~\cite{Hryczuk:2011vi}. TC is supported by DoE contract number DE-AC02-76SF00515.    ML is supported by the Simons Postdoctoral Fellows Program.  AP is supported by DoE grant DE-SC0007859 and CAREER grant NSF-PHY 0743315.  TRS is supported by the NSF under grants PHY-0907744 and AST-0807444, and by the U.S. DoE under cooperative research agreement DE-FG02-05ER41360.  This research was supported in part by NSF grant NSF PHY11-25915.   TC thanks the Galileo Galilei Institute and the INFN for partial support during the completion of this work.  AP thanks UC Berkeley, where this work was initiated. 

\appendix
\section{Computing the Sommerfeld Enhancement}
\label{Appendix: Sommerfelding}
To determine the relic density of the wino today, as well as the flux of its annihilation products, we must properly account for the SE.  When a pair of non-relativistic neutralinos/charginos annihilate, they experience a potential due to some combination of Yukawa and Coulomb interactions arising from the exchange of gauge bosons in ladder diagrams.  In the non-relativistic limit of the potential, and for the $l=0$ partial wave, \emph{i.e.}, $s$-wave annihilation, the two-body reduced wavefunction $\psi(x)$ is given by the time-independent Schr{\"o}dinger equation
\begin{equation}
\psi''(x) = \Bigg( \frac{V(x)}{E} - 1 \Bigg)\, \psi(x),
\label{eq:dimlessschrodinger}
\end{equation}
where $E = p^2/m_\chi$ and $x = p\,r$ for neutralino mass $m_\chi$. Note that $E$ is always defined as the energy relative to the $\chi^0 \,\chi^0$ state at zero velocity:
\be
p = \frac{m_\chi}{2}\sqrt{v^2 + 4\,\sum_i \frac{\delta_i}{m_{\chi}}},
\ee
where $v$ is the physical relative velocity between the two particles, as in the main text, and $\delta_i = m_i - m_\chi$.
$V(x)$ arises from the long-range interactions from gauge boson exchange and assumes the non-relativistic limit.  

The possible two-body $s$-wave states can be described by the magnitude of the total charge $Q$ and total spin $S$.  For the wino system, there are five distinct possibilities, $(Q, S) = (0,0), (0, 1), (1, 0), (1, 1), (2, 0)$.  The last four correspond to single-state systems, where only a single two-body state participates in the interaction.  Their potentials take the form 
\begin{equation}\label{eq: Generic potential}
V(r) = \Delta - \frac{a}{r} - \frac{b}{r} e^{-m_A r},
\end{equation}
where $m_A$ is the relevant vector boson mass and $\Delta = 2 \,\delta$ for the $Q=2$ and $(Q,S)=(0,1)$ states and $\Delta=\delta$ for the $Q=1$ states ($\delta$ is the mass splitting between the neutralino and chargino).  Depending on the incoming states, $m_A$ could be $m_Z$ or $m_W$.  The Coulomb (Yukawa) potential has a coefficient $a$ ($b$).  Note that positive $a$ and $b$ correspond to attractive potentials.  For concreteness, $a$, $b$, and $m_A$ for all the one-state wino processes are given in Table ~\ref{tab: single state a and b} \cite{Hisano:2006nn}.  Note that  $c_w\equiv \cos \theta_W = \sqrt{1-s_w^2}$, and $\alpha_W$ is the weak coupling.

\begin{table}[h!]
\renewcommand{\arraystretch}{1.5}
\setlength{\tabcolsep}{7pt}
\begin{tabular}{|c|ccc|}
\hline
(Q,S) & a & b & $m_A$\\ 
\hline
(0,1) & $\alpha$ & $\alpha_W\,c_w^2$ & $m_Z$ \\
(1,0) & 0 & $\alpha_W$ & $m_W$\\
(1,1) & 0 & $\alpha_W$ & $m_W$\\
(2,0)  & $-\alpha$ & $-\alpha_W\,c_w^2$ & $m_Z$\\
\hline

\hline
\end{tabular}
\caption{The values of $a$, $b$, and $m_A$ for the wino model \cite{Hisano:2006nn}.}
\label{tab: single state a and b}
\end{table}

The $(Q,S) = (0, 0)$ system is a two-state system in which the $\chi^0\, \chi^0$ and $\chi^+ \, \chi^-$ two-body states are coupled.  In this case, the potential is a $2\times2$ matrix, where the off-diagonal elements describe the couplings between these states:

\begin{equation} 
\frac{V(x)}{E} = \left( \begin{matrix} 0 & \quad\quad-\sqrt{2} \left(\frac{\alpha_W m_\chi}{x \,p} \right) e^{-\frac{m_W \,x}{p}} \\ 
\vspace{-10pt}\\
-\sqrt{2} \left(\frac{\alpha_W m_\chi}{x\,p} \right) e^{-\frac{m_W \,x}{p}} &\quad\quad \frac{2 \, m_\chi \, \delta}{p^2} - \frac{\alpha_W m_\chi s_w^2}{x\,p} - \left(\frac{\alpha_W m_\chi c_w^2}{x\,p} \right) e^{-\frac{m_Z \,x}{p}}   \end{matrix} \right) \, .
\end{equation}

The SE is obtained by solving \eref{eq:dimlessschrodinger} for the appropriate choice of \eref{eq: Generic potential} and boundary conditions. For the one-state systems, the boundary conditions are $\psi'(x) \rightarrow i \, k\, \psi(x)$ so that $\psi(x) \sim e^{i \,k \,x}$ is purely outgoing as $x\rightarrow \infty$, and $\psi(0) = 1$.  Note that these are \emph{not} the physical boundary conditions for the reduced wavefunction, but lead to a particularly simple expression for the SE.   Here, the dimensionless momentum $k = \sqrt{1 - \delta/E}$.  When $E < \delta$, the two-body initial state is not on-shell because $E$ is always defined relative to the $\chi^0\, \chi^0$ state, and the appropriate boundary condition is instead that the wavefunction is exponentially falling as $x \rightarrow \infty$. 

Given a solution $\psi$, the SE for the one-state system is
\begin{equation}
s = \psi(\infty) \quad\quad\quad \left(\text{one-state system}\right),
\end{equation}
and the enhanced annihilation cross section is
\begin{equation}
\sigma \,v = c \,\Gamma\, |s|^2 \quad\quad\quad \left(\text{one-state system}\right),
\end{equation}
where $\Gamma$ is the perturbative annihilation cross section for the two-body system, and $c = 2 \, (1)$ for annihilation of identical (distinct) particles.  For the explicit $s$-wave zero-velocity expressions for $\Gamma$ in the wino model, see \cite{Hisano:2006nn}.  We dress the cross sections for wino annihilations to $W^+\,W^-$ and $\gamma\,Z^0$ final states in the present day with the appropriate propagator and phase-space factors, which are important for $M_2 \sim 100 \GeV$.

In an $n$-state system, the wavefunction is an $n$-vector $\psi_i(r) \mbox{ with }i = 1,\ldots,n,$ and the Schr{\"o}dinger equation must be solved with $n$ different sets of boundary conditions.  In all cases, the boundary condition $\psi_i(\infty)$ is a purely outgoing wave (for states above threshold) or is exponentially falling (for states below threshold), and $\psi_i(0) = \delta_{ij}$, $j = 1,\dots,n$.  For the wino system, the only coupled case is the simplest one, $n=2$ and the $i$ index labels different two-body states: $\chi^0\,\chi^0$, $\chi^+\, \chi^-$. For clarity we will often label the states by their particle content, \emph{e.g.}~writing $i = 00$ for $\chi^0\, \chi^0$. The large-$x$ values of these $n$ solutions yield $n$ $n$-vectors, which are used to build up the SE matrix,
\begin{equation} 
s_{ij} = \psi_i(\infty) \quad\quad\quad \left(n\text{-state system}\right),
\end{equation}
and the enhanced annihilation cross section is
\begin{equation}
\label{eq:xsectionfromsommerfeld}
\sigma_i \,v = c_i \,\sum_{j, j'} s_{ij} \, \textbf{$\Gamma$}_{jj'} \,s^*_{ij'}  \quad\quad\quad \left(n\text{-state system}\right),
\end{equation}
where the ``hard annihilation matrix" $\Gamma_{jj'} \equiv \int \mathcal{A}_j^\dag\,\mathcal{A}_{j'}$ and the integral is over Lorentz-invariant phase space.

In principle, numerically solving the Schr{\"o}dinger equation is straightforward.  In practice, matching onto an oscillating solution at infinity is more numerically challenging than matching onto a constant one.  Furthermore, the infinite range of the Coulomb potential means that when such a term is present in $V(x)$, one needs to be careful to check for convergence, \emph{i.e.}, by integrating out to large enough $x$ such that the solution has entered a regime where the solution approximates $e^{i k x}$. However, the exact solution of the one-state Schr{\"o}dinger equation for a Coulomb potential is known analytically.  We take the general approach of factoring out the known solutions, plane wave or Coulomb as is appropriate, and solving for their coefficients.

In the one-state system, we rewrite the solution to the Schr{\"o}dinger equation as $\psi(x) = \xi(x)\, \phi(x)$, where $\phi(x)$ is the Coulomb/plane wave solution with appropriate boundary conditions: $\phi(0) = 1$ and $\phi(x)$ purely outgoing as $x \rightarrow \infty$. Given these boundary conditions, $\xi(0) = 1$, and $\xi(x)$ should approach a constant value at large $x$, or $\xi'(x) \rightarrow 0$.

First, consider the single-state case, where the Schr{\"o}dinger equation takes the form
\begin{equation}
 \psi''(x) = \left(- \frac{a\, m_\chi}{x\,p} - \frac{b\, m_\chi}{x \, p} \,e^{-\frac{m_A x}{p}} + \frac{m_\chi \, \Delta}{p^2} - 1 \right) \psi(x).
  \end{equation}
Requiring that $\phi''(x) =  \left(- \frac{a\,m_\chi}{x\,p} + \frac{m_\chi \, \Delta }{p^2} - 1 \right) \phi(x)$, and writing $\psi(x) = \xi(x)\, \phi(x)$, the equation for $\xi(x)$ becomes,
\begin{align}\xi''(x) + 2 \left( \frac{\text{d}}{\text{d}x} \ln \phi(x) \right) \xi'(x)  & = \left( - \frac{b\, m_\chi}{x\,p} \,e^{-\frac{m_A x}{p}} \right) \xi(x). \end{align}

We can analytically compute the coefficient $ \frac{\text{d}}{\text{d}x} \ln \phi(x)$ on the LHS, and then solve the differential equation for $\xi(x)$ with the boundary conditions described above.  In this simple case, the SE for the annihilation rate is given by $s = |\phi(\infty)\, \xi(\infty)|^2 = S_\mathrm{coulomb} |\xi(\infty)|^2$. The Sommerfeld enhancement for a Coulomb potential is known analytically and given by
\begin{equation} S_\mathrm{coulomb} = \frac{\pi\, \alpha \, m_\chi}{p} \frac{1}{1 - e^{-\pi\, \alpha\, m_\chi / p}}. \end{equation}

If there is no Coulomb term ($a=0$), then $\phi(x) = e^{i\, \sqrt{1 - \Delta / E}}$ for all $E > \Delta$ (for $E$ below $\Delta$ the two-body state does not represent real scattering particles). The coefficient $ \frac{\text{d}}{\text{d}x} \ln \phi(x)$ is now trivial, but the calculation otherwise proceeds as above.

A similar approach can be taken for the $(Q,S)=(0,0)$ system where the $\chi^0 \,\chi^0$ and $\chi^+ \,\chi^-$ states are coupled, with the latter experiencing a Coulomb interaction. Writing the two-state solution as $\big( \phi_1(x)\, \xi_1(x), \, \phi_2(x)\, \xi_2(x) \big)$, and using the potential matrix  given above, we can define $\phi_1(x) = e^{ix}$, and $\phi_2(x)$ to satisfy $\phi_2''(x) = \left(\frac{2\, m_\chi \delta}{p^2} - 1 - \left(\frac{\alpha_w m_\chi s_w^2}{p} \right) \frac{1}{x}\right) \phi_2(x)$.  Then, the differential equation for $\big(\xi_1(x), \xi_2(x)\big)$ becomes

\bea
&&\left( \begin{matrix} \xi_1''(x) + 2 i \xi_1'(x) \\ \xi_2''(x) + 2 \left( \frac{d}{dx} \ln \phi_2(x) \right) \xi_2'(x) \end{matrix} \right) \nonumber\\
&&\quad= - \left( \begin{matrix} 0 & \sqrt{2} \left(\frac{\alpha_w m_\chi}{x\,p} \right) e^{-\frac{m_W x}{p}} \frac{\phi_2(x)}{\phi_1(x)} \\[1em]
 \sqrt{2} \left(\frac{\alpha_w m_\chi}{x\,p} \right) e^{-\frac{m_W x}{p}} \frac{\phi_1(x)}{\phi_2(x)} &  \left(\frac{\alpha_w m_\chi c_w^2 }{x\,p} \right) e^{-\frac{m_Z x}{p}}  \end{matrix} \right) \left( \begin{matrix} \xi_1(x) \\ \xi_2(x) \end{matrix} \right). 
\eea

There is a problem with this approach when the chargino state $\chi^+ \chi^-$ is below threshold; namely, the Coulomb solution has zeroes so there may be no finite $\xi_2(x)$ such that $\psi_2(x) = \xi_2(x)\, \phi_2(x)$. However, when the chargino state is below threshold, we are only interested in its effects on the $\chi^0\, \chi^0$ annihilation, and all such effects are suppressed by a factor of $e^{-(m_Z/p) x}$. Thus, the solution converges quickly outside the range of the $Z^0$-mediated Yukawa potential even though the virtual $\chi^+ \,\chi^-$ state has a Coulomb interaction.

Given this argument, below threshold we define $\phi_2(x) = e^{- x \sqrt{2 \,m_\chi \delta /p^2 - 1}}$. This leads to
\begin{align} 
& \left( \begin{matrix} \xi_1''(x) + 2 i \xi_1'(x) \\[1em]
 \xi_2''(x) - \left(2 \,\sqrt{\frac{2\, m_\chi \delta}{p^2} - 1}\right) \xi_2'(x) \end{matrix} \right) = \nonumber \\
&  - \left( \begin{matrix} 0 & \sqrt{2} \left(\frac{\alpha_w m_\chi}{x\,p} \right) e^{-\left[ i + \frac{m_W}{p} + \sqrt{\frac{2 m_\chi \delta}{p^2} - 1} \right] x} \\[1em]
\sqrt{2} \left(\frac{\alpha_w m_\chi}{x\,p} \right) e^{-\left[-i + \frac{m_W}{p} - \sqrt{\frac{2 m_\chi \delta}{p^2} - 1} \right] x}  & \frac{\alpha_w m_\chi s_w^2}{x\,p} +  \left(\frac{\alpha_w m_\chi c_w^2}{x\,p} \right) e^{-\frac{m_Z \,x}{p}}  \end{matrix} \right) \left( \begin{matrix} \xi_1(x) \\ \xi_2(x) \end{matrix} \right). 
\end{align}

One of the terms in this matrix has the potential to be exponentially \emph{enhanced}, with a coefficient of the form $e^{\left[\sqrt{2 m_\chi \delta/p^2 - 1} - m_W/p\right] x}$. However, for this term to act as an enhancement rather than a suppression, we would need to have $\sqrt{2 m_\chi \delta/p^2} > m_W/p \Leftrightarrow m_\chi > m_W^2/(2\, \delta)$. For $m_W \simeq 80$ GeV and $\delta \simeq 0.17$ GeV, this corresponds to $\sim 19$ TeV DM. Consequently, for the parameter space of interest, there is no convergence issue. For $m_\chi \lesssim 5$ TeV, this modification increases the range of the potential (relative to the usual Yukawa potential from $m_W$) by less than a factor of 2.

As in the one-state case, the appropriate boundary condition at infinity is that $\xi_i'(x) \rightarrow 0$. At the origin, we require that $\xi_i(0) = \psi_i(0)$, and solve the Schr{\"o}dinger equation twice with $\big(\xi_1(0), \xi_2(0)\big) = (1,0)$ and $\big(\xi_1(0), \xi_2(0)\big) = (0,1)$ as described above. The entries in the Sommerfeld matrix are given by
\begin{equation} 
s_{ij} = \xi_i(\infty) \phi_i(\infty).
\end{equation}
As usual, the $s_{2i}$ matrix elements vanish below threshold.

\section{Subtraction Procedure}
\label{Appendix: Subtraction}

\begin{figure}[t!]
\centering
\includegraphics[width=0.8\textwidth]{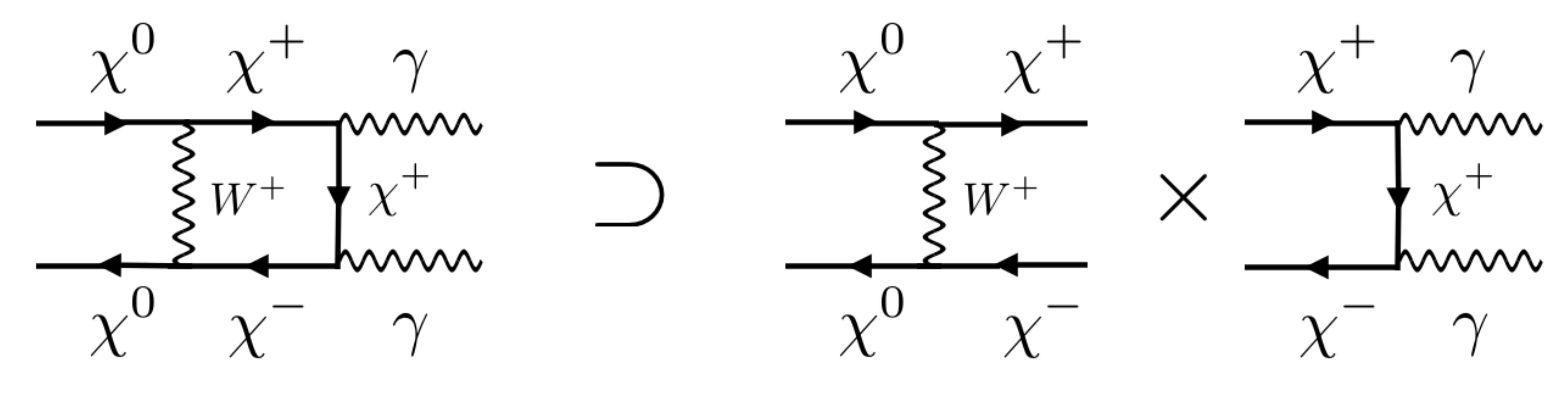}
\caption{
The Feynman diagram for the process $\chi^0\,\chi^0 \rightarrow \gamma \,\gamma$ includes a piece that is generated when applying the SE to the tree-level chargino annihilation to photons.  When including the leading 1-loop corrections to the hard cross section, care must be taken to not include this quantity twice.
}
\label{Fig: Ann to Photons Diagram}
\end{figure}

Working at tree-level for the hard matrix element plus the SE, \emph{i.e.}, tree-level-SE, can give rise to a non-zero cross section for a process whose tree-level amplitude for annihilation is zero.  For concreteness, consider the annihilation process $\chi^0 \chi^0\rightarrow \gamma \gamma$.  At tree-level, $\mathcal{A}^{\gamma \gamma}_{00} = 0$; the only non-zero term in the $2 \times 2$ annihilation sub-matrix for this final state is $\Gamma^{\gamma \gamma}_{+ -,+ -}$.   A non-zero rate for $\chi^0 \chi^0 \rightarrow \gamma \gamma$ is generated when the SE is included because the $|s_{00,+-}|^2 \,\Gamma^{\gamma \gamma}_{+-,+-}$ term is non-zero as shown in Fig.~\ref{Fig: Ann to Photons Diagram}.
(The same structure applies for annihilation to $\gamma Z$ and $ZZ$.) 

Working to the next order in perturbation theory, \emph{i.e.}, one-loop-SE,  the annihilation matrix gains additional terms at higher order in the gauge coupling $g$:
\begin{equation}\label{Eq: Gamma^GammaGamma}
\Gamma^{\gamma \gamma} = \bigintsss \left( \begin{matrix} 0 & \mathcal{A}^{\gamma\gamma\, *}_{00}\big(g^4\big) \,\mathcal{A}^{\gamma \gamma}_{+-}\big(g^2\big)  \\  
\vspace{-10 pt}\\
\mathcal{A}^{\gamma\gamma}_{00}\big(g^4\big) \,\mathcal{A}^{\gamma \gamma\,*}_{+-}\big(g^2\big)&\quad\quad \Big |\mathcal{A}^{\gamma \gamma}_{+-}\big(g^2\big)  \Big |^2 + 2 \,\mathrm{Re} \Big[\mathcal{A}^{\gamma \gamma\,*}_{+-}\big(g^2\big)\,\mathcal{A}^{\gamma\gamma}_{+-}\big(g^4\big)\Big] \end{matrix} \right) + \OO\big(g^8\big). 
\end{equation}
where the $g^n$ in each parenthesis denotes the lowest order in perturbation theory where the amplitude is non-zero. 

It is important to avoid double counting the one-loop contribution to the SE which is also contained in the $\OO\big(g^4\big)$ amplitudes.  The rest of this appendix is devoted to understanding the subtraction procedure of \cite{Hryczuk:2011vi}.  To begin, it is necessary to compute the leading piece of the unresummed SE.  This can be found by taking the low-velocity limit, where only the $\chi^0 \chi^0$ initial state is relevant:
\begin{equation}
s_{00,+-} =  \sqrt{2}\, \alpha_W \,\frac{M_2}{m_W}; \quad\quad \quad s_{00,00} = 1,
\end{equation}
where $\alpha_W \frac{M_2}{m_W} \ll 1$ has been assumed.  For larger values of $M_2$, this approximation breaks down and the SE must be obtained numerically (see Appendix~\ref{Appendix: Sommerfelding} for more details).  Following \cite{Hryczuk:2011vi}, we define $W_\text{step}$ as the contribution to the SE from each additional rung of the SE ladder diagram, given schematically as
\be\label{eq: s leading}
s_{00,+-} \sim \sum_{n=1}^\infty \left( \prod_{i=1}^n W_\text{step}^i\right).
\ee
At one-loop order, $W_\text{step} \equiv \sqrt{2}\,  \alpha_W \,\frac{M_2}{m_W}$, from \eref{eq: s leading}.  Then (continuing to use annihilation to photons as an explicit example), Fig.~\ref{Fig: Ann to Photons Diagram} can be translated into
\be\label{Eq: Subtraction Amplitude}
\mathcal{A}_{00}^{\gamma \gamma}\big(g^4\big) \supset W_\text{step}\big(g^2\big) \times \mathcal{A}_{+-}^{\gamma \gamma}\big(g^2\big).
\ee
This is exactly the quantity that must be subtracted from the hard amplitude to avoid double counting in Eqs.~(\ref{Eq: Gamma^GammaGamma}) and (\ref{Eq: Sigma^GammaGamma}).

This subtraction completely removes the leading contribution to $\mathcal{A}^{\gamma\gamma}_{00}$ at high masses, which scales as $\alpha\, \alpha_W / m_W$. To see that this is the case, it is useful to take the large $M_2$ limit where analytic expressions can be utilized.  The tree-level perturbative cross section for chargino annihilation into photons is
\be
\sigma_{+-,+-}^{\gamma \gamma} \,v= \frac{\pi \,\alpha^2}{M_2^2},
\ee
and the unsubtracted 1-loop perturbative cross section for neutralino annihilation into photons is \cite{Bern:1997ng}
\be
\Big[\sigma_{00,00}^{\gamma \gamma}\,v\Big]_\text{perturbative} = \frac{4\, \pi\,\alpha^2 \,\alpha_W^2}{m_W^2} \quad \quad \quad\text{for } M_2 \rightarrow \infty.
\ee

The corresponding terms in the annihilation matrix $\Gamma$ are $\frac{\pi \,\alpha^2}{M_2^2}$ and $ \frac{2\, \pi\,\alpha^2 \,\alpha_W^2}{m_W^2} $ respectively, taking the $c_i$ factors of \eref{eq:xsectionfromsommerfeld} into account. Noting that the corresponding amplitude for both of these processes are real, an analytic estimate for the subtracted cross section is 
\begin{eqnarray}
\Big[\sigma_{00,00}^{\gamma \gamma}\,v\Big]_\text{subtracted} & =& 2 \int \left| s_{00,+-} \mathcal{A}^{\gamma \gamma}_{+-} + s_{00,00} \mathcal{A}^{\gamma \gamma}_{00} \right|^2 \nonumber \\
&\underset{ M_2 \rightarrow \infty}{\xrightarrow{\quad\quad\quad}}& 2 \left( \sqrt{2}\, \frac{\alpha_W M_2}{m_W} \times \sqrt{\pi} \frac{\alpha}{M_2} - \sqrt{2\, \pi}\, \frac{\alpha\, \alpha_W}{m_W} \right)^2  \nonumber \\
&\underset{ M_2 \rightarrow \infty}{\xrightarrow{\quad\quad\quad}}& 0.
\end{eqnarray}
This demonstrates that at large masses the leading $(1/m_W)^2$ piece of the 1-loop annihilation cross section is entirely captured by the Sommerfeld enhancement, which resums this contribution and preserves unitarity (see \emph{e.g.} \cite{Hisano:2004ds} for a discussion). The subtracted amplitude does not have any terms that scale as $1/m_W$.

\pagebreak

\begin{spacing}{1.1}
\bibliography{ExcludingWinosBib}
\bibliographystyle{utphys}
\end{spacing}
\end{document}